\documentclass[11pt]{article}
\usepackage{amsmath,amssymb}
\usepackage[nosort]{cite}
\usepackage[margin=1in]{geometry}
\usepackage{epsf}
\usepackage{graphicx}
\makeatletter \@addtoreset{equation}{section} \makeatother

\newcommand{\fft}[2]{{\frac{#1}{#2}}}
\newcommand{\ft}[2]{{\textstyle\frac{#1}{#2}}}
\def\nn{\nonumber}

\let\bm=\bibitem

\newcommand{\be}{\begin{equation}}
\newcommand{\ee}{\end{equation}}
\def\ba{\begin{array}}
\def\ea{\end{array}}
\def\ft#1#2{{\textstyle{\frac{\scriptstyle #1}{\scriptstyle #2}}}}
\def\fft#1#2{\frac{#1}{#2}}
\def\del{\partial}

\def\sst#1{{\scriptscriptstyle #1}}

\def\dalemb#1#2{{\vbox{\hrule height .#2pt
        \hbox{\vrule width.#2pt height#1pt \kern#1pt
                \vrule width.#2pt}
        \hrule height.#2pt}}}
\def\square{\mathord{\dalemb{6.8}{7}\hbox{\hskip1pt}}}

\newcommand{\hoch}[1]{$\, ^{#1}$}
\newcommand{\bea}{\begin{eqnarray}}
\newcommand{\eea}{\end{eqnarray}}

\newcommand{\Tr}{{\rm Tr} }
\def\0{{\sst{(0)}}}
\def\1{{\sst{(1)}}}
\def\2{{\sst{(2)}}}
\def\3{{\sst{(3)}}}
\def\4{{\sst{(4)}}}
\def\5{{\sst{(5)}}}
\def\6{{\sst{(6)}}}
\def\7{{\sst{(7)}}}
\def\8{{\sst{(8)}}}

\def\R{\rlap{\rm I}\mkern3mu{\rm R}}

\def\R{\rlap{\rm I}\mkern3mu{\rm R}}

\def\R{{{\mathbb R}}}

\begin{document}
\begin{flushright}
MCTP-07-12\ \ \ \ MIFP-07-08\\
{\bf hep-th/0703184}\\
March\  2007
\end{flushright}

\vspace{10pt}
\begin{center}

{\Large {\bf Bubbling AdS Black Holes}}

\vspace{20pt}

James T. Liu$^{\dagger}$, H. L\"u$^{\ddagger}$, C.N. Pope$^{\ddagger}$
and Justin F. V\'azquez-Poritz$^{\ddagger}$

\vspace{20pt}

{\hoch{\dagger}\it Michigan Center for Theoretical Physics\\
Randall Laboratory of Physics, The University of Michigan\\
Ann Arbor, MI 48109-1040, USA}

\vspace{10pt}

{\hoch{\ddagger}\it George P. \&  Cynthia W. Mitchell Institute
for Fundamental Physics\\
Texas A\&M University, College Station, TX 77843-4242, USA}

\vspace{40pt}

\underline{ABSTRACT}
\end{center}

We explore the non-BPS analog of `AdS bubbles', which are regular
spherically symmetric 1/2 BPS geometries in type IIB supergravity.
They have regular horizons and can be thought of as bubbling
generalizations of non-extremal AdS black hole solutions in
five-dimensional gauged supergravity. Due to the appearance of the
Heun equation even at the linearized level, various approximation and
numerical methods are needed in order to extract information about
this system. We study how the vacuum expectation value and mass of a
particular dimension two chiral primary operator depend on the
temperature and chemical potential of the thermal Yang-Mills
theory. In addition, the mass of the bubbling AdS black holes is
computed. As is shown numerically, there are also non-BPS solitonic
bubbles which are completely regular and arise from continuous
deformations of BPS AdS bubbles.

\newpage
\tableofcontents
\addtocontents{toc}{\protect\setcounter{tocdepth}{2}}

%%%%%%%%%%%%%%%%%%%%%%%%%%%%%%%%%%%%%%%%
\section{Introduction}

Black holes in five dimensions have been extensively studied in the
framework of both ungauged and gauged supergravity theories.
Furthermore, many of these explicit studies have been performed in the
context of the STU model, which corresponds to $\mathcal N=2$
supergravity coupled to two vector multiplets. Because of the presence
of the graviphoton and two vector fields, the STU model generally
admits three-charge black holes with up to two rotation parameters.
In fact, in the ungauged context, many solutions have been
constructed, and their explicit forms have often been proven to be
useful, especially in the context of black hole thermodynamics and
stringy microstate counting.

At the same time, AdS black holes (and their variants) in gauged
supergravity theories have found widespread application in the study of
the AdS/CFT correspondence. For instance, an $R$-charged black hole in
global AdS$_5$ geometry corresponds to equilibrium non-zero
temperature ${\cal N}=4$ SU($N$) supersymmetric Yang-Mills theory on
$S^3\times \R$ with finite chemical potential. Five-dimensional BPS
`black holes' were first constructed in \cite{Behrndt:1998ns},
although there it was also realized that they are in fact naked
singularities. Unlike the ungauged case, angular momentum must be
turned on in order to obtain true BPS black holes in gauged
supergravity.  This was done in \cite{Gutowski:2004ez,Gutowski:2004yv}
for one rotation parameter, and subsequently generalized to two
independent rotations. As in the ungauged case, these black holes
admit non-extremal generalizations \cite{Behrndt:1998jd}, which can be
further generalized to include rotations
\cite{Cvetic:2004hs,Cvetic:2004ny,Chong:2005da,Kunduri:2006ek,Chong:2006zx}.

BPS objects play an important r\^ole in AdS/CFT, regardless of their
precise nature, since they necessarily survive in both the strong and
weak coupling regimes of the duality. Along these lines, the BPS naked
singularities were shown in \cite{Myers:2001aq} to correspond to a
distribution of giant gravitons in AdS$_5\times S^5$, where they were
also denoted `superstars.' In terms of the Lin, Lunin and Maldacena
(LLM) boundary conditions for 1/2 BPS configurations
\cite{Lin:2004nb}, these superstars correspond to a disk in the phase
space of free fermions with a uniform shade of gray, which may be
interpreted as a coarse graining of an underlying spacetime foam
picture of gravity \cite{Balasubramanian:2005mg}.

The investigation of smooth 1/2 BPS geometries in \cite{Lin:2004nb}
also led to the construction of a new set of 1/2 BPS `AdS bubbles'
which eliminate the naked singularity of the singular black holes-- not
through angular momentum but rather by turning on additional scalar
fields $\varphi_i$. These scalars are present in the full $\mathcal
N=8$ gauged supergravity but lie outside of the $\mathcal N=2$
truncation \cite{Lin:2004nb,Chong:2004ce}.  These AdS bubbles are in
fact completely regular geometries without horizons and, as such, admit
a description in the LLM language as a deformation of the AdS disk
into an ellipse in the phase space of free fermions.

In order to generalize the above five-dimensional single-charge AdS
bubbles to include three charges, the authors of \cite{Chong:2004ce}
considered a consistent truncation of five-dimensional $\mathcal N=8$
gauged supergravity that retains the three U(1) gauge fields in the
maximal torus of the SO(6) gauge group along with five scalar fields.
Although this truncated system is not itself the bosonic sector of any
supersymmetric theory, it nevertheless allows the construction and
lifting of BPS solutions to yield supersymmetric configurations of the
full $\mathcal N=8$ theory and hence the original IIB supergravity as
well.

Although LLM configurations are by their very nature pure states, and
not thermal ones, it is natural to investigate the effect of turning
on non-zero temperature starting from a particular LLM background.  The
main goal of this paper is to initiate such an exploration by constructing
AdS bubbles away from the BPS limit.  In particular, we will look for
non-extremal AdS black holes with regular horizon and carrying $\varphi_i$
scalar deformations of the same sort encountered in the BPS bubbles of
\cite{Lin:2004nb,Chong:2004ce}.  We will refer to these solutions as
`bubbling AdS black holes,' which are a subset of thermal AdS bubbles.
Though not the focus of our paper, we will demonstrate numerically that
there are also non-BPS solitonic AdS bubbles which are completely regular
and horizon-free.

We note that a coarse-graining of LLM geometries was considered in
\cite{buchel}, and expanded upon in
\cite{Balasubramanian:2005mg,Giombi:2005zq}, where the free fermion
configuration was given an equilibrium non-zero temperature.  Using
the LLM correspondence, this configuration can then be mapped into a
`hyperstar' supergravity background by appropriate transformation of
the Fermi-Dirac distribution into the LLM $z(x_1,x_2,y)$ function.  It
is not clear, however, whether this background actually describes a
non-zero temperature field theory; it is by no means obvious that
simply giving the matrix eigenvalues (the `free fermions') a
Fermi-Dirac profile would correspond to turning on non-zero
temperature in $\mathcal N=4$ super Yang-Mills.  If this were the
case, then one would expect the corresponding supergravity background
to have a horizon and to break supersymmetry. In \cite{buchel}, it was
pointed out that the coarse-graining has been taken over only the
half-BPS sector of the full Hilbert space of type IIB supergravity,
and that the result of using the complete Hilbert space should be a
non-BPS background that has a horizon.  Because of these
considerations, we shall not pursue this direction of coarse-graining,
but will look directly for bubbling AdS black holes in the
supergravity itself.

For the case of the AdS black hole, the main effect of moving away
from the extremal limit is to introduce a `renormalization' of the
charge. One might hope that the AdS bubble might be generalized away
from the BPS limit in a similar way, in which much of the original
structure survives in a `renormalized' form. Unfortunately, upon
closer inspection of the equations of motion, it does not appear that
the AdS bubble can be made non-extremal in such a simple manner. Thus,
we must content ourselves with either approximate or numerical
methods.

We will proceed by performing a linearized analysis of the
second-order equations of motion. Although the non-extremal AdS bubble
solution is not known, explicit solutions are known in the two
separate limits of either turning off non-extremality or turning off
the $\varphi_i$ deformation.  Since we will focus on solutions for
which there is a horizon, we choose to consider the initial background
to be that of the non-extremal $R$-charged black hole. We will turn on
the additional scalars $\varphi_i$ at linear order, which will then
backreact onto the other fields at higher order.

The linearized $\varphi_i$ equations are second-order equations with
four regular singular points, and hence may be mapped to the Heun
equation.  Unfortunately, in contrast with the hypergeometric equation
which has three regular singular points, much less is known about the
solutions to the Heun equation. In particular, the general two-point
connection problem, {\it i.e.} relating local solutions at two regular
singular points, remains unsolved.

This is of course not the first time where the Heun equation has
arisen in the study of the wave equation in AdS black hole
backgrounds. In particular, calculations involving quasi-normal modes
for black holes have generally led to expressions related to the Heun
equation. In such cases, various methods have been applied in order to
obtain approximate solutions of the resulting system.  A common
method, also used in black hole absorption calculations, involves
matching approximate solutions in two overlapping regions: an inner
region containing the horizon and an outer region that includes
asymptotic infinity.  Provided that there is a large overlap, then
essentially complete information may be reliably extracted from this
approach. Higher-order corrections can be included via a perturbative
approach. We will implement this approximation technique for the
regime $T\gg\mu_i$, where $T$ and $\mu_i$ are the temperature
and chemical potentials of the field theory,
respectively. Furthermore, there is a high temperature limit
$T\gg 1$, along with $T\gg\mu_i$, for which there is an
approximate solution which covers the entire region from
the horizon to asymptotic infinity without the need for matching.

The asymptotic behavior of $\varphi_i$ is related to perturbations
away from the UV superconformal fixed point of the dual field
theory. In particular, as in the BPS case, the normalizable mode of
$\varphi_i$ corresponds to giving a vacuum expectation value (vev) to
the dimension two chiral primary operator $\Tr\,Z_i^2$, where
$Z_i=\fft{1}{\sqrt{2}}(\phi_1^i+{\rm i}\phi_2^i)$ and $\phi_1^i$ and
$\phi_2^i$ are three pairs of real scalars of ${\cal N}=4$ super
Yang-Mills theory, for $i=1,2,3$. On the other hand, the
non-normalizable mode of $\varphi_i$ corresponds to a term of the form
$\Tr\,Z_i^2$ in the Lagrangian of the conformal field theory. This
massive term is only present for nonzero temperature, which could be
indicative of a phase transition in the field theory at zero
temperature. We will discuss how these field theory deformations
depend on the physical parameters of the thermal field theory, namely
the temperature, chemical potential and $R$-charge, in both the grand
canonical ensemble and the canonical ensemble.

The paper is organized as follows. In section 2, we review the
previously-known AdS black hole and BPS bubble solutions. In section
3, we perform a linearized analysis of the bubbling AdS black hole. In
particular, we consider the linearized $\varphi_i$ equations in the
background of the AdS black hole.  Focusing on the single-charge case,
matching and perturbation techniques are used to find approximate
solutions to the linearized $\varphi_1$ equation. Properties of the
$\varphi_1$ solution are discussed. Specifically, the asymptotic
behavior is matched with perturbations of the dual field theory as
functions of temperature and chemical potential. In section 4, we
consider the backreaction of $\varphi_1$ onto the metric and other
matter fields, and in section 5 we then discuss the mass of these
non-BPS bubbles.  In section 6, we present some numerical support for
the existence of bubbling AdS black holes for arbitrary values of
$\varphi_1$.  Lastly, we conclude in section 7.

%%%%%%%%%%%%%%%%%%%%%%%%%%%%%%%%%%%%%%%%
\section{Review of AdS black holes and BPS bubbles}

\subsection{AdS black hole}

The bosonic Lagrangian for the STU model takes the form
\begin{equation}
\mathcal L=R*\mathbf1-\ft12\sum_{\alpha=1}^2 *d\phi_\alpha\wedge
d\phi_\alpha -\ft12\sum_{i=1}^3X_i^{-2}{*F}^i\wedge F^i-V{*\mathbf1}
+F^1\wedge F^2\wedge A^3,
\end{equation}
where the $X_i$'s are constrained scalars satisfying $X_1X_2X_3=1$, and
which may be taken to be
\begin{equation}
X_1=e^{-\fft1{\sqrt6}\phi_1-\fft1{\sqrt2}\phi_2},\qquad
X_2=e^{-\fft1{\sqrt6}\phi_1+\fft1{\sqrt2}\phi_2},\qquad
X_3=e^{\fft2{\sqrt6}\phi_1}.
\label{eq:x1x2x3}
\end{equation}
The scalar potential is given by
\begin{equation}
V=-4g^2\sum_{i=1}^3X_i^{-1}=-4g^2(X_2X_3+X_3X_1+X_1X_2),
\end{equation}
and the ungauged system is recovered by setting $g=0$.

The three-charge AdS$_5$ black hole solution is given by
\begin{eqnarray}
ds^2 &=& -\mathcal H^{-2/3}fdt^2+\mathcal
H^{1/3}(f^{-1}dr^2+r^2d\Omega_3^2),
\nonumber\\
A_{(1)}^i &=& -\coth\beta_iH_i^{-1}dt,\qquad X_i=\mathcal
H^{1/3}H_i^{-1},\qquad
\varphi_i=0,\nonumber\\
f &=& 1-\fft{m}{r^2}+g^2r^2\mathcal H,\qquad \mathcal H=H_1H_2H_3,
\label{eq:blackh}
\end{eqnarray}
where the harmonic functions are given by
\begin{equation}
H_i=1+\fft{q_i}{r^2},\qquad q_i\equiv m \sinh^2\beta_i.
\label{eq:neharm}
\end{equation}
We shall focus primarily on the case of a single charge $q_1$, for
which the roots of $f$ are given by
%%%%
\be r_{\pm}^2=-\fft{1}{2g^2} (1+g^2 q_1)\pm \fft{1}{2g^2} \sqrt{(1+g^2
q_1)^2+4g^2 m}\,.
\ee
%%%%
The event horizon is located at $r_h=r_+$. Notice that, in the
one-charge case with $q_1>0$, any positive value of $m$ guarantees a
regular horizon. On the other hand, for the two and three-charge
cases, a regular horizon may always be obtained for sufficiently large
$m$.

This supergravity background is dual to equilibrium non-zero
temperature ${\cal N}=4$ SU($N$) supersymmetric Yang-Mills theory on
$S^3\times \R$ with chemical potentials for the $U(1)$ $R$-charges.
The temperature of the field theory is equated with the Hawking
temperature of the black hole, which is \cite{cveticgubser}
%%%%
\be
T=\fft{2r_h^6+r_h^4(1+\sum_i q_i)-\prod_i q_i}{2\pi r_h^2 \prod_i
\sqrt{r_h^2+q_i}}\,,\label{temp}
\ee
%%%%
where the horizon radius $r_h$ is the largest root of $f$. We have set
$g=1$ for simplicity. Likewise, the $R$-charge chemical potentials
$\mu_i$ of the field theory are equated with the electric potentials
at the horizon, which are
%%%%
\be
\mu_i=\fft{Q_i}{r_h^2+q_i}\,,\label{chempot}
\ee
%%%%
where
%%%%
\be
Q_i^2=q_i(r_h^2+q_i)\Big[ 1+\fft{1}{r_h^2} \prod_{j\ne i}
(r_h^2+q_j)\Big]\,.\label{Qi}
\ee
%%%%
Also, the physical charges $Q_i$ of the AdS black hole correspond to
$R$-charges in the dual field theory. These relations will be useful
for expressing various results in terms of the physical quantities of
the field theory.

\subsection{BPS AdS bubble}

The authors of \cite{Chong:2004ce} considered a consistent truncation
of five-dimensional $\mathcal N=8$ gauged supergravity retaining the
three U(1) gauge fields in the maximal torus of the SO(6) gauge group
along with five scalar fields. In $\mathcal N=2$ language, this
corresponds to taking the bosonic sector of the STU model and coupling
it to three additional scalars $\varphi_i$, which are not described by
special geometry.  The Lagrangian is given by \cite{Chong:2004ce}
\begin{eqnarray}
\mathcal L&=&R*\mathbf1-\ft12\sum_{\alpha=1}^2 
*d\phi_\alpha\wedge d\phi_\alpha
-\ft12\sum_{i=1}^3X_i^{-2}*{F}^i\wedge F^i-V*{\mathbf1}
+F^1\wedge F^2\wedge A^3\nonumber\\
&&-\ft12\sum_{i=1}^3 *d\varphi_i\wedge d\varphi_i
-2g^2\sum_{i=1}^3\sinh^2\varphi_i {*A}^i\wedge A^i,
\label{eq:lagnew}
\end{eqnarray}
where the scalar potential has the modified form
\begin{equation}
V=2g^2\Bigl(\sum_{i=1}^3X_1^2\sinh^2\varphi_i
-2\sum_{i<j}^3X_iX_j\cosh\varphi_i\cosh\varphi_j\Bigr).
\label{eq:spotnew}
\end{equation}
The original STU model scalars satisfy the same constraints as above,
and in particular may be given by (\ref{eq:x1x2x3}).

In order to ensure a supersymmetric solution, the regular three-charge
AdS bubble configuration then takes the form \cite{Chong:2004ce}
\begin{eqnarray}
ds^2 &=& -\mathcal H^{-2/3}f\,dt^2+\mathcal H^{1/3}
(f^{-1}dr^2+r^2d\Omega_3^2),
\nonumber\\
A_{(1)}^i &=& -H_i^{-1}dt,\qquad X_i=\mathcal H^{1/3}H_i^{-1},\qquad
\cosh\varphi_i=(xH_i)',\nonumber\\
f &=& 1+g^2x\mathcal H,\qquad\mathcal H=H_1H_2H_3,
\label{eq:adsbubble}
\end{eqnarray}
where $x\equiv r^2$, and a prime denotes a derivative with respect to
$x$.  The above solution is fully determined up to the functions
$H_i$, which must satisfy the conditions
\begin{equation}
f(xH_i)''=-g^2[(xH_i)'^2-1]\mathcal HH_i^{-1},
\label{eq:susyeom}
\end{equation}
to ensure that the equations of motion are satisfied.

For the one-charge case, corresponding to a 1/2 BPS configuration, we
may take $H_2=H_3=1$. Then the equation of motion
(\ref{eq:susyeom}) reduces to
\begin{equation}
[(1+g^2 xH_1)^2]''=2g^4,
\end{equation}
which admits a general solution of the form
\begin{equation}
H_1=\sqrt{1+\fft{2(1+g^2q_1)}{g^2x}+\fft{\overline c^2}{g^4x^2}}
-\fft1{g^2x}.\label{eq:bubsol}
\end{equation}
Here $q_1$ is the $R$-charge, and $\overline c$ is a constant related to the
$\varphi_1$ scalar deformation.  Note that, at large distances, $H_1$ admits
the expansion
\begin{equation}
H_1\sim 1+\fft{q_1}{x}+\cdots,
\end{equation}
while regularity of the AdS bubble at short distances demands
$\overline c=1$.  In addition, the BPS naked singularity of
\cite{Behrndt:1998ns,Behrndt:1998jd} is recovered by taking $\overline
c=1+g^2 q_1$, in which case $H_1$ reduces to the standard `harmonic
function' form $H_1=1+q_1/x$.

In general, as we will demonstrate below, the scalar $\varphi_1$ carries
$E_0=2$, and hence has an asymptotic expansion of the form
\begin{equation}
\varphi_1\sim\fft{c_1+c_2\log x}x+\cdots,
\end{equation}
However, based on the explicit solution (\ref{eq:bubsol}), we see that
the log term vanishes, $c_2=0$, while
%%%%
\be
c_1 =\sqrt{(1+q_1)^2-\overline c^2}=\sqrt{q_1(q_1+2)}\,,
\label{normalization}
\ee
%%%%
where the second expression is for the regular AdS bubble solution. We
have set $g=1$.  Applying the AdS/CFT dictionary (see {\it
e.g.}~\cite{review}), this implies that the dimension two chiral
primary operator $\Tr\,Z_1^2$ gets a vev $c_1$, where
$Z_1=\fft{1}{\sqrt{2}}(\phi_1+{\rm i}\phi_2)$ and $\phi_1$ and
$\phi_2$ are two of the six real scalars of ${\cal N}=4$ super
Yang-Mills theory.

Although we are not aware of any closed form expressions, it can be
seen that regular two and three-charge solutions to (\ref{eq:susyeom})
also exist.  These solutions correspond to 1/4 and 1/8 BPS AdS
bubbles, and can be realized in the framework of LLM configurations
with fewer supersymmetries
\cite{Donos:2006iy,Donos:2006ms,Kim:2005ez}.

%%%%%%%%%%%%%%%%%%%%%%%%%%%%%%%%%%%%%%%%
\section{Linearized analysis of the bubbling AdS black hole}

Given the relatively simple non-extremal generalization
(\ref{eq:blackh}) of the BPS AdS black hole solution, we have been led
to look for a corresponding non-extremal version of the AdS bubble
(\ref{eq:adsbubble}) where the $\varphi_i$ scalars are present.  In
the absence of supersymmetry, we no longer have the benefit of working
with first-order Killing spinor equations.  However, the form of the
non-extremal black hole (\ref{eq:blackh}) is curiously close to that
of the BPS limit; the primary difference is that the charge parameters
$q_i$ in the harmonic functions $H_i$ are `renormalized' as follows:
$q_i\to m \sinh^2\beta_i$.  This suggests that perhaps a similarly
straightforward generalization may be obtained for the AdS bubble,
where much of the BPS structure might survive, except perhaps in
`renormalized' form.

Unfortunately, closer inspection of the equations of motion arising
from the Lagrangian (\ref{eq:lagnew}) does not suggest any simple
manner in which the AdS bubble solution may be made non-extremal.  In
particular, retaining the BPS-like relation $\cosh\varphi_i =(xH_i)'$
leads to either the possibility that $\varphi_i=0$, in which case
$H_i$ takes the form (\ref{eq:neharm}), or to a constrained system of
equations which only appear to admit a natural solution of the BPS
bubble form given by (\ref{eq:susyeom}).  As a result, we must use
either approximate or numerical methods when moving away from
extremality.

\subsection{Linearized $\varphi_i$ equations}

In this section, we explore the basic features of the non-extremal
AdS bubble by performing a linearized analysis of the second-order
equations of motion obtained from (\ref{eq:lagnew}).  We proceed by
noting that, although it is not clear how to write down the
complete solution for a non-extremal AdS bubble, explicit forms for the
solutions are known in the two separate limits of either turning off
non-extremality or turning off the $\varphi_i$ deformation.  Since
we are mainly interested in solutions with a horizon, we choose to
start from the non-extremal $R$-charged black hole background of 
(\ref{eq:blackh}) and (\ref{eq:neharm}) and then turn on the additional
scalars $\varphi_i$ at linear order.  Turning on these scalars will
then backreact onto the other fields.  However this backreaction occurs
at the next order, and may be ignored in the initial analysis.  (We
will return to the backreaction in another section.)

At linearized order we are only concerned with the linearized
equation of motion for $\varphi_i$, which takes the form
\begin{equation}
\Bigl[\square-4g^2X_i(2X_i-\sum_{j=1}^3X_j)-4g^2(A_\mu^i)^2\Bigr]
\varphi_i=0.
\label{eq:varphieom}
\end{equation}
Here the scalars $X_i$ take on the background values given above in
(\ref{eq:blackh}) and (\ref{eq:neharm}).  There is a slight subtlety
for the background gauge fields, however, related to the `gauge fixed'
form of the action (\ref{eq:lagnew}).  In particular, while the
standard black hole solution (\ref{eq:blackh}) is invariant under
gauge transformations of $A^i$, the above equation of motion is not.
Taking this into account, we allow a constant term in the expression
for the background electric potential
\begin{equation}
A^i=(b^i-\coth\beta_iH_i^{-1})dt\equiv A_t^i\, dt.
\label{eq:shifta1}
\end{equation}
We shall see below that this constant must be chosen to make the
potential vanish at the horizon.  The motivation for this requirement
can already be seen by noting that the invariant square of the
electric potential, $(A_\mu^i)^2$, which acts as a source in
(\ref{eq:varphieom}), blows up at the horizon unless $A_t^i$ is
arranged to vanish there.

By substituting the background fields into (\ref{eq:varphieom}), we
obtain the scalar equation
\begin{eqnarray}
&&(x^2f\varphi_i')'+g^2\biggl[\sum_{j\ne i}^3(x+q_j)+(x^2f)^{-1}
\prod_{j\ne i}^3(x+q_j)\Bigl(-g^2\prod_{j\ne i}^3(x+q_j)\nonumber\\
&&\kern16em
+\fft{m x}{q_i}
-2b^i\sqrt{1+m/q_i}\,x+(b^i)^2(x+q_i)\Bigr)\biggr]\varphi_i =0,
\label{eq:seom}
\end{eqnarray}
where
\begin{equation}
x^2f=g^2\prod_{j=1}^3(x+q_j)+x^2-m x,
\end{equation}
is a cubic polynomial, and where $q_i\equiv m \sinh^2\beta_i$.  Note that,
in the event $q_i=0$, the above equation is replaced by the considerably
simpler expression
\begin{equation}
(x^2f\varphi_i')'+\fft{g^2}x\Bigl(x^2-\prod_{j\ne i}^3q_j\Bigr)\varphi_i=0,
\end{equation}
which may be obtained by directly taking $A^i=0$ in (\ref{eq:varphieom}).

Since $x^2f$ is cubic, it can be seen that (\ref{eq:seom}) is a
second-order equation with three regular singular points at the roots
of $x^2f$. Including the singular point at infinity, which is also regular, this
equation in fact has four regular singular points, and hence may be
mapped into the general Heun equation.  Unfortunately, the general
two-point connection problem is as yet unsolved. However, general
features of the linearized $\varphi_i$ deformation may be extracted
from the second-order equation (\ref{eq:seom}).

Recalling that the goal of the linearized analysis is to turn on a
$\varphi_i$ deformation starting from the $R$-charged black hole
background, we demand that the solution to (\ref{eq:seom}) be regular
and bounded in the entire region from the horizon to the spatial
boundary at infinity.  Before examining the solution, we find it
convenient to trade the non-extremality parameter $m$ with the horizon
location $x_h$, defined to be the largest positive root of $f(x)$.
Note that, in terms of $x_h$, we have
\begin{equation}
m=x_h+\fft{g^2}{x_h}\prod_{i=1}^3(x_h+q_i),
\label{eq:muexp}
\end{equation}
along with the factorization
\begin{equation}
x^2 f=g^2(x-x_h)(x^2+Q x-R),
\end{equation}
where
\begin{equation}
Q=x_h+\fft1{g^2}+q_1+q_2+q_3,\qquad
R=\fft{q_1q_2q_3}{x_h}.
\end{equation}
Examination of the indicial equation around the horizon $x_h$ (which
is a regular singular point) yields the characteristic exponents $\pm\zeta_i$
where
\begin{equation}
\zeta_i^2=-\fft{\prod_{j=1}^3(x_h+q_j)}{(x_h^2+Qx_h-R)^2}
\left(b^i-\sqrt{1+\fft{m}{q_i}}\fft{x_h}{x_h+q_i}\right)^2.
\end{equation}
Noting that the expression above is non-positive (since $x_h+q_j$ must
be positive to avoid naked singularities), we immediately see that the
characteristic exponents are purely imaginary, except for the case
when they vanish.  Since imaginary exponents give rise to undesirable
oscillatory solutions of the form $\varphi_i\sim
\sin(|\zeta_i|\log(x-x_h))$, we conclude that the constant $b^i$ in
the electric potential must be adjusted to satisfy
\begin{equation}
b^i=\sqrt{1+\fft{m}{q_i}}\fft{x_h}{x_h+q_i}=\fft{\coth\beta_i}{H_i(x_h)}.
\label{eq:bireq}
\end{equation}
This ensures that the potentials given by (\ref{eq:shifta1}) indeed
vanish at the horizon, $A^i(x_h)=0$, thus confirming what we
had alluded to above.

Demanding that the electric potentials vanish at the horizon, the
$\varphi_i$ equation can now be put into the form
\begin{eqnarray}
&&((x-x_h)(x^2+Q x-R)\varphi_i')'+\biggl[\sum_{j\ne i}^3(x+q_j)
-(x^2+Qx-R)^{-1}\prod_{j\ne i}^3(x+q_j)\nonumber\\
&&\kern4em\times\biggl(\fft{x_h^2+Qx_h-R}{x_h+q_i}
+(x-x_h)\biggr)\biggr]\varphi_i=0.
\end{eqnarray}
The solutions to this equation may be characterized by the Riemann
$P$-symbol
\begin{equation}
P\left\{\begin{matrix}
x_h&x_1&x_2&\infty\\
0&\alpha_1&\alpha_2&1&;x\\
0&-\alpha_1&-\alpha_2&1
\end{matrix}\right\},
\end{equation}
where $x_1$ and $x_2$ are the two roots of the quadratic 
equation $x^2+Qx-R=0$.
Their associated exponents, $\alpha_1$ and $\alpha_2$, may easily 
be obtained, although
we have no particular need for their explicit forms.  We do note, however,
that the characteristic exponents sum to $2$, which is in agreement with the
general theory of the Heun equation.

\bigskip
\noindent\underline{The single-charge case}
\bigskip

Since we are primarily interested in the non-extremal generalization
of the single-charge 1/2 BPS bubble, we now focus on the case where
$q_2=q_3=0$.  In this case, we assume the corresponding scalars
$\varphi_2$ and $\varphi_3$ ought not to be excited.  As a result, we
are left with a single equation for $\varphi_1$, which now takes the
form
\begin{equation}
(x(x-x_h)(x+Q)\varphi_1')'+x\left[1+\fft{x_h+Q}{x+Q}\fft{q_1}{x_h+q_1}
\right]\varphi_1=0.
\end{equation}
This equation still contains four regular singular points, and as such
can be brought into Heun form.  In particular, we may map the singular
points $\{0,x_h,-Q\}$ to $\{0,1,a\}$ by introducing
\begin{equation}
x=zx_h,\qquad Q=-ax_h,
\end{equation}
after which the scalar equation becomes
\begin{equation}
(z(z-1)(z-a)\varphi_1')'+z\left[1+\zeta\fft{1-a}{z-a}\right]\varphi_1=0.
\label{eq:z1eqn}
\end{equation}
Here the prime denotes a derivative with respect to $z$.
We have also defined the dimensionless parameter
\begin{equation}
\zeta=\fft{q_1}{x_h+q_1}.
\label{eq:zetdef}
\end{equation}
Note that $\zeta\to1$ in the limit of large charge, while $\zeta\to0$
in the limit of vanishing charge.

The scalar equation (\ref{eq:z1eqn}) can be brought into the canonical
form of the Heun equation through the substitution
$\varphi_1=(z-a)^{\pm\sqrt\zeta} \tilde\varphi_1$.  As a result, the
solution may be written in terms of a local Heun function
\begin{equation}
\varphi_1=\fft{1-a}{z-a}Hl(\fft1a,-\fft\zeta{a};1-\sqrt\zeta,1+
\sqrt\zeta,1,1;\fft{z-1}{z-a}).
\label{eq:vp1loc}
\end{equation}
Here we have imposed the boundary condition that $\varphi_1$ is
regular at the horizon. In particular, the expansion
of the local Heun function gives
\begin{equation}
\varphi_1=1+(1+\zeta)\fft{z-1}{a-1}+\fft{4-a+(6-a)\zeta+\zeta^2}4\fft{(z-1)^2}
{(a-1)^2}+\cdots.
\label{eq:horizex}
\end{equation}
where $\zeta$ is given in (\ref{eq:zetdef}), and where
\begin{equation}
a=-\left(1+\fft{q_1+1/g^2}{x_h}\right).
\label{eq:adef}
\end{equation}
In principle, we would also like to obtain the expansion of
$\varphi_1$ in the asymptotic regime $z\to\infty$ in order to extract
the boundary behavior (\ref{eq:vphias}).  However, the two-point
connection problem for the Heun equation is in general a difficult
task, and there is as yet no straightforward way to connect the
behavior of (\ref{eq:vp1loc}) near the horizon with (\ref{eq:vphias})
near the boundary. In order to match the horizon and boundary
behaviors of $\varphi_1$, we need to utilize further approximation
techniques. Before turning to this, we shall first consider the
general behavior of $\varphi_i$.

\bigskip
\noindent\underline{General behavior}
\bigskip

The asymptotic behavior of the solution for $\varphi_i$ is governed by
the point at infinity.  In particular, given the repeated characteristic
exponent of $1$, it may be shown that the solution near the boundary
has the form
\begin{equation}
\varphi_i(x)\sim\fft{c_1^i+c_2^i\log x}{x}+\cdots,\qquad x\to\infty,
\label{eq:vphias}
\end{equation}
which is consistent with $\varphi_i$ being an $E_0=2$ scalar (or
equivalently being associated with conformal dimension $\Delta=2$ in
the dual gauge theory).  Although the two independent solutions
encoded in (\ref{eq:vphias}) admit different interpretations in the
dual gauge theory, they are both allowed to be present in the
supergravity solution.

In particular, as in BPS case, $c_1^i$ can be interpreted as the vev
of the dimension 2 chiral primary operator $\Tr\,Z_i^2$, where
$Z_i=\ft{1}{\sqrt{2}}(\phi_1^i+{\rm i}\phi_2^i)$ and $\phi_1^i$ and
$\phi_2^i$ are three pairs of real scalars of ${\cal N}=4$ super
Yang-Mills theory, for $i=1,2,3$. On the other hand, the $c_2^i$ term
corresponds to adding the relevant deformation $c_2^i\,\Tr\, Z_i^2$ to
the Lagrangian of the conformal field theory. The $c_2^i$ term is only
present when one moves away from extremality, which indicates that
there may be a phase transition in the field theory at zero
temperature.

Turning to the horizon, we see that, with the choice of $b^i$ in
(\ref{eq:bireq}), the repeated characteristic exponent of $0$ at
$x=x_h$ indicates that the solution near the horizon has the form
\begin{equation}
\varphi_i(x)\sim d_1^i + d_2^i\log(x-x_h)+\cdots,\qquad x\to x_h.
\label{eq:vphiho}
\end{equation}
Clearly, we must set $d_2^i=0$ to avoid a logarithmic divergence of
$\varphi_i$ at the horizon.  

In principle, this boundary condition
now fixes the complete solution, in the sense that the coefficients
$c_1^i$ and $c_2^i$ in the asymptotic expansion may be determined
directly from $d_1^i$ and the physical parameters of the background
solution.  However, in the absence of any general connection formulae
for the Heun equation, there is no straightforward way to make this
relation explicit.  (We note that, even if the generic connection
matrix were known, it might not be applicable to this solution as it
has repeated characteristic exponents.) In order to match the horizon 
and boundary behaviors of $\varphi_1$ for the single-charge case, we 
instead turn to approximate solutions of equation (\ref{eq:z1eqn}).

\subsection{Matching approximate solutions}

We first consider an approximation technique which involves matching
approximate solutions in two overlapping regions. In particular, one
region contains the horizon while the other includes asymptotic
infinity.  Provided these two regions overlap, the solutions may then
be matched up in the overlap region.  A drawback of this approach is
that it is not always possible to ensure a large overlap region,
depending on the physical parameters of the system.  However, if such
a large overlap exists, then essentially complete information may be
reliably extracted using this matching.

Noting that $a$ introduces a new scale into the problem, we may
consider solving (\ref{eq:z1eqn}) in the two regions: $i$) the
asymptotic region where $z\gg1$, and $ii$) the horizon region where
$z\ll|a|$.  Provided $|a|\gg1$, these two regions will have a large
overlap ($1\ll z\ll|a|$) where reliable matching may be performed.
From (\ref{eq:adef}), we see that overlap is ensured for either
$q_1\gg x_h$ or $x_h\ll 1/g^2$.

\subsubsection{The asymptotic region}

To highlight the asymptotic region, $z\gg1$, we may rewrite the scalar
equation (\ref{eq:z1eqn}) as
\begin{equation}
\varphi_1''+\left(\fft2w+\fft1{w-1}\right)\varphi_1'
+\fft1{w(w-1)}\left(1-\fft\zeta{w-1}\right)\varphi_1
=-\fft1a\left[\fft1{w(w-1/a)}\left(\varphi_1'+\fft{1+\zeta}{w-1}\varphi_1
\right)\right],
\label{eq:asexp}
\end{equation}
where we have introduced the rescaled coordinate $w=z/a$.  In this case,
the additional factor of $1/a$ on the right-hand side of (\ref{eq:asexp})
allows us formally to develop a solution for $\varphi_1$ as a perturbative
expansion in $1/a$
\begin{equation}
\varphi_1=\varphi^{(0)}+\fft1a\varphi^{(1)}+\fft1{a^2}\varphi^{(2)}\cdots,
\end{equation}
where $\varphi^{(0)}$ solves the homogeneous equation corresponding to
the left-hand side of (\ref{eq:asexp}).  Since this can be put into
hypergeometric form, the solution is essentially known.  In practice,
however, matching of the asymptotic and horizon expansions is
facilitated by introducing yet another expansion, this time in
$\zeta$.  Examination of (\ref{eq:zetdef}) indicates that there are
two relevant limits to consider, namely the large and small charge
limits.

The large charge limit corresponds to $q_1\gg x_h$, or equivalently
$\zeta\approx1$.  In this case, we let $\zeta=1+\hat\zeta$ and rearrange
(\ref{eq:asexp}) to read
\begin{equation}
\varphi_1''+\left(\fft2w+\fft1{w-1}\right)\varphi_1'
+\fft{w-2}{w(w-1)^2}\varphi_1=\fft{\hat\zeta}{w(w-1)^2}\varphi_1
-\fft1a\left[\fft1{w(w-1/a)}\left(\varphi_1'+\fft{2+\hat\zeta}{w-1}
\varphi_1\right)\right].
\label{eq:asexp1}
\end{equation}
Solutions to this equation can now be developed as a double expansion
in $\hat\zeta$ and $1/a$.  Although this may seem to be only a slight
rearrangement of (\ref{eq:asexp}), the main simplification here is that
the homogeneous equation can now be solved in terms of elementary
functions, $\varphi_1^{(0)}=c_1u_1+c_2u_2$ where
\begin{equation}
u_1=\fft1{w-1},\qquad u_2=\fft1{w-1}\left(\log(-w)+\fft1w\right).
\end{equation}
At each successive order in the perturbation, the lower order solutions
feed in as sources on the right-hand side of (\ref{eq:asexp1}).  However,
since the homogeneous solutions are elementary, the inhomogeneous system
has a straightforward solution which can be developed, {\it e.g} through
variation of parameters.

Up to first order in both $\hat\zeta$ and $1/a$, we find that the two
linearly independent solutions for the outside function can be expressed
as
\begin{eqnarray}
\varphi_{\rm out}^1&=&\fft1{w-1}\left[1-\fft{\hat\zeta}2
\left(\log\Bigl(1-\fft1w\Bigr)
-\fft1{w}\right)-\fft1{aw}\biggl(1-\fft{\hat\zeta}2
\Bigl(\log\Bigl(1-\fft1w\Bigr)-2+\fft1{2w}\Bigr)\biggr)
+\cdots\right],\nonumber\\
\varphi_{\rm out}^2&=&\fft1{w-1}\Biggl[\log(-w)+\fft1w
-\fft{\hat\zeta}2\biggl(\log(-w)\Bigl(\log\Bigl(1-\fft1w\Bigr)-\fft1w\Bigr)
+\Bigl(2-\fft1w\Bigr)\log\Bigl(1-\fft1w\Bigr)\nonumber\\
&&\kern3em+\log^2\Bigl(1-\fft1w\Bigr)+2{\rm Li}_2\Bigl
(\fft1{1-w}\Bigr)\biggr)
+\fft1{aw}\left(-3-\log(-w)+\fft1{2w}+\mathcal O(\hat\zeta)\right)
+\cdots\Biggr],\nonumber\\
\label{eq:outexp}
\end{eqnarray}
where we recall that $w=z/a$.  We have organized these two solutions
according to the asymptotic behavior
\begin{equation}
\varphi_{\rm out}^1\sim \fft{a}z+\cdots,\qquad
\varphi_{\rm out}^2\sim \fft{a}z\log z+\cdots,
\end{equation}
corresponding to the normalizable and non-normalizable $E_0=2$ modes
of $\varphi_1$, as in (\ref{eq:vphias}).

\subsubsection{The horizon region}

Turning next to the horizon region, $z\ll|a|$, we now choose to write
the scalar equation (\ref{eq:z1eqn}) as
\begin{equation}
\varphi_1''+\left(\fft1z+\fft1{z-1}\right)\varphi_1'
=\fft1a\left(\fft1{1-z/a}\right)\left[\varphi_1'+\fft1{z-1}
\left(1+\zeta\fft{1-1/a}{1-z/a}\right)\varphi_1\right].
\end{equation}
This again allows us to develop an expansion in $1/a$, where the
independent solutions to the homogeneous equation on the left-hand
side are simply
\begin{equation}
u_1=1,\qquad u_2=\log\fft{z-1}z.
\end{equation}
These two solutions correspond directly to the near horizon behavior
given in (\ref{eq:vphiho}); in particular, we see that only $u_1$ is
well behaved at the horizon.  Developing this solution to the first
few orders in $1/a$ gives
\begin{eqnarray}
\varphi_{\rm in}&=&1+\fft1{2a}(1+\zeta)(z-1+\log z)+
\fft1{24a^2}\Bigl[(z-1)\bigl(9(1-\zeta^2)+2(\zeta^2+7\zeta+4)z\bigr)
\nonumber\\
&&-\bigl(5\zeta^2+14\zeta+5-6(1+\zeta)^2z\bigr)\log z
+3(1+\zeta)^2\log^2z+6(1+\zeta)^2{\rm Li}_2\Bigl(1-\fft1z\Bigr)\Bigr]
+\cdots.\nonumber\\
\label{eq:inexp}
\end{eqnarray}
Note that this is purely an expansion in $1/a$, and in particular it
is valid for arbitrary $\zeta$.  Furthermore, it can be seen that this
expression for $\varphi_{\rm in}$ agrees with the near horizon
expansion given in (\ref{eq:horizex}) in the overlapping region of
validity $z\to1$ and $a\to\infty$.  The advantage of (\ref{eq:inexp})
over (\ref{eq:horizex}), however, is that here $\varphi_{\rm in}$
remains valid even for $z$ away from the horizon (provided $z\ll|a|$).
This is precisely what is needed in order to match the horizon
expression with the asymptotic forms of the solution given above in
(\ref{eq:outexp}).

\subsubsection{Matching}

While the asymptotic and horizon solutions (\ref{eq:outexp}) and
(\ref{eq:inexp}) were derived under the independent conditions of
$z\gg1$ and $z\ll|a|$, they are both valid in the overlap region $1\ll
z\ll|a|$, so long as $|a|\gg1$.  In order to match the solutions in
this overlap region, it is convenient to rewrite the horizon solution
(\ref{eq:inexp}) in terms of $w$.

After suitable rearrangement, $\varphi_{\rm in}$ then takes the form
of a series solution in $w$ as well as an expansion in $1/a$
\begin{eqnarray}
\varphi_{\rm in}&=&\left[1+\fft{w}2(1+\zeta)+
\fft{w^2}{12}(\zeta^2+7\zeta+4)+\mathcal O(w^3)\right]\nonumber\\
&&+\fft1a\left[\fft12(1+\zeta)(\log(w a)-1)+
\fft{w}{24}(-11\zeta^2-14\zeta+1+6(1+\zeta)^2\log(wa))+
\mathcal O(w^2)\right]\nonumber\\
&&+\mathcal O\Bigl(\fft1{a^2}\Bigr).
\label{eq:horser}
\end{eqnarray}
Here it is important to realize that, although the expressions are
no longer complete at each new order in $1/a$, the resulting series
in $w$ are in principle well behaved for $|w|<1$.  This is what allows
a consistent matching with the asymptotic solution (\ref{eq:outexp}),
with $\varphi_{\rm out}$ similarly expanded as a series in $w$.

The asymptotic solution to the scalar equation is in general a
linear combination of the two solutions given in (\ref{eq:outexp}):
\begin{equation}
\varphi_{\rm out}(x)=c_1\varphi_{\rm out}^1+c_2\varphi_{\rm out}^2
\sim\fft{ax_h(c_1+c_2\log(-x/ax_h))}x,
\label{eq:asympx}
\end{equation}
where we have transformed back to the coordinate $x=r^2$.
By matching this with the normalized horizon solution (\ref{eq:horser}),
we find
\begin{eqnarray}
c_1&=&-1+\fft1a\biggl(1-\log(-a)-\fft{\hat\zeta}2
\Bigl(\fft{\pi^2}3+1+\log(-a)\Bigr)\biggr)+\mathcal O(\hat\zeta^2)
+\mathcal O\Bigl(\fft1{a^2}\Bigr),\nonumber\\
c_2&=&\fft{\hat\zeta}2+\fft1a\biggl(-1+\fft{\hat\zeta}2\log(-a)\biggr)
+\mathcal O(\hat\zeta^2)+\mathcal O\Bigl(\fft1{a^2}\Bigr),
\end{eqnarray}
where 
%%%
\be
\hat\zeta=-\fft{x_h}{x_h+q_1}\,,
\ee
%%%
and $a$ is given by (\ref{eq:adef}). Note that both $\hat\zeta$ and
$a$ are negative for physical values of the charge and horizon radius.

In order to more readily apply the AdS/CFT dictionary for examining
the asymptotic behavior, we express (\ref{eq:asympx}) in terms of
the original $r$ coordinate:
%%%%
\be
\varphi_{\rm out}(r)=\fft{ax_h[c_1-c_2 \log(-ax_h)]}{r^2}+2ax_h c_2
\, \fft{\log r}{r^2}\,.
\ee
%%%%
Note that we can always include factors of $g$ to ensure that the
logarithms have dimensionless arguments. However, we have set
$g=1$ for simplicity.  The above $1/r^2$ term corresponds to giving a
vev to the chiral primary operator $\Tr\,Z^2$. The
$(\log r)/r^2$ term, on the other hand, corresponds to adding a
massive deformation $-\fft12 {\bar m}^2\, \Tr\, Z^2$ to the Lagrangian
of the conformal field theory. {\it A priori}, there is an ambiguity
in the normalization of $\varphi_1$. Recall that, for the BPS AdS
bubble, the thermal mass ${\bar m}=0$ while, from (\ref{normalization}),
$\Tr\,Z^2$ gets a vev $v_1\equiv\sqrt{q_1(q_1+2)}$.  Therefore, since
we are interested in a thermalization of $\mathcal N=4$ super-Yang
Mills on top of the 1/2 BPS sector specified by a given vev $v_1$, we
may fix the normalization by taking $\langle \Tr\,Z^2\rangle=v_1$ for
the bubbling AdS black hole. Then, taking the limit $q_1\gg x_h$, we
find
%%%%
\be
{\bar m}^2\approx 4x_h\Big( \fft{1}{q_1+1}-\fft{1}{2q_1}\Big)
\sqrt{q_1(q_1+2)}\,.\label{m2}
\ee
%%%%
As $x_h$ vanishes, the thermal mass goes to zero. This is certainly
not unexpected, since the thermal mass vanishes in the BPS limit,
though at the same time it would not have been surprising if ${\bar
m}$ did not behave in a smooth manner in this limit. Interestingly
enough, ${\bar m}$ also vanishes for $q_1\rightarrow 1$. As we will see, this could be due to a phase transition in the field theory that is associated with the Hawking-Page transition. In
particular, for $q_1<1$, the operator $\Tr\,Z^2$ could become tachyonic,
thereby destabilizing the moduli space of the field theory.

We would now like to express ${\bar m}$ in terms of the physical
parameters of the thermal field theory, namely the temperature $T$,
chemical potential $\mu_1$ and $R$-charge $Q_1$. These quantities are
all expressed in terms of the AdS black hole parameters $q_i$ and
$x_h=r_h^2$ in (\ref{temp}), (\ref{chempot}) and (\ref{Qi}). In the
limit $q_1\gg x_h$ for a single charge,
%%%%
\be
T=\fft{1+q_1}{2\pi\sqrt{q_1}}\,,\qquad \mu_1=\sqrt{1+x_h}\,,
\qquad Q_1=q_1 \sqrt{1+x_h}\,.
\ee
%%%%
We therefore have a choice of expressing $x_h$ and $q_1$ in terms of
two out of the three physical parameters of the field theory, which
corresponds to different ensembles. For example, the $R$-charge $Q_1$
is held fixed in the canonical ensemble. Therefore, the thermal mass
is a function of temperature and chemical potential in the grand
canonical ensemble, while it depends on temperature and $R$-charge in
the canonical ensemble.

We first work in the grand canonical ensemble. We can express $x_h$
and $q_1$ in terms of $T$ and $\mu_1$ in the limit $q_1\gg x_h$ as $x_h
\approx \mu_1^2-1$ and $q_1 \approx 2\pi T(\pi T\pm \sqrt{\pi^2
T^2-1})-1$. Notice that black holes with two different values of $q_1$
correspond to the same temperature.  However, since the entropy is
given by $S\approx2\pi x_h\sqrt{q_1}$, the black hole with larger
$q_1$ is entropically favorable and so we shall take the $+$ sign in
the expression for $q_1$. Since $x_h \ge 0$, in this regime
$\mu_1^2\ge 1$. Also, $\pi T>1$ in order for $q_1$ to be real.

One way in which $q_1\gg x_h$ is if $T\gg\mu_1$. This implies that
$T\gg 1$ and $q_1\approx 4\pi^2 T^2$. We can then express the thermal
mass as ${\bar m}^2\approx 2(\mu_1^2-1)$. Thus, for a given chemical
potential, the thermal mass vanishes at zero temperature and
approaches a constant in this high temperature limit, which may imply
a sort of saturation taking place.  Also, the thermal mass gets
enhanced by increased chemical potential. Notice that this high
temperature regime does not include the point at which the thermal
mass vanishes at a finite temperature.

Alternatively, $q_1\gg x_h$ can be satisfied without having to take
$T$ to be large by taking $\mu_1\approx 1$. Then after expressing the
thermal mass in (\ref{m2}) in terms of $T$ and $\mu_1$, we find that
${\bar m}=0$ for $\pi T=1$, which corresponds to $q_1=1$. As $T$ increases, the thermal mass asymptotically
approaches the constant value discussed in the previous large
temperature limit, that is ${\bar m}^2\rightarrow 2(\mu_1^2-1)$. 

We will now look at the situation in the canonical ensemble, for which the
$R$-charge $Q_1$ is held fixed. One way to satisfy $q_1\gg x_h$ is to
consider $T^3\gg Q_1$ and $T\gg 1$. Then we have
%%%%
\be
{\bar m}^2\approx 2 \Big( \fft{Q_1^2}{16\pi^4 T^4}-1\Big)\,,
\ee
%%%%
where ${\bar m}^2\ge 0$ since $x_h\ge 0$. We see that the thermal mass
decreases with temperature and increases with the $R$-charge $Q_1$. Note that in order for there to be a horizon at $x_h>0$, ${\bar m}^2>0$. This requires that $Q_1>4\pi^2 T^2$.

We will now briefly discuss the conditions for the regime $q_1\gg x_h$
to be consistent with local thermodynamical stability constraints and
dynamical considerations for the charged AdS black hole that were
given in \cite{cveticgubser}, in both the canonical ensemble and grand
canonical ensemble. At the level of linearized $\varphi_1$, we do not
need to consider the backreaction on this background, which would then
alter these thermodynamical relations.

In the grand canonical ensemble, in order for the AdS black hole to be
dynamically preferred over pure AdS, $\pi T>\pi T_c\approx 1$, where
$T_c$ is the temperature of the Hawking-Page transition. Furthermore,
the local thermodynamical stability constraint is satisfied only for
$\pi T\approx 1$. Thus, we only have a small
window for which the above calculations are consistent with
stability. The temperature at which ${\bar m}$ vanishes
lies within this window; in particular, this occurs at the Hawking-Page transition and presumably signifies the corresponding phase transition in the field theory. On the other hand, the large temperature regime where ${\bar m}$ saturates does not lie within this window.
For the canonical ensemble, the local stability constraint is satisfied for $\pi
T\approx\sqrt{2}$, which also satisfies the condition for the AdS
black hole to be dynamically preferred over pure AdS.

We would now like to recall the approximations that have been made,
namely that $|\hat\zeta|\ll1$ and $|a|\gg1$ correspond to the large
charge limit $q_1\gg x_h$, which is equivalent to $T\gg\mu_1$. We
could also match approximate solutions in the small charge limit
$q_1\ll x_h$, which means that $|\zeta|\ll 1$. Then in order to be
consistent with the condition that the asymptotic and horizon regions
have a large overlap, we require that $|a|\gg 1$, which further
implies that $x_h\ll 1$. We can use (\ref{temp}) and (\ref{chempot})
to express these conditions as $\mu_1\ll 1\ll T$. Therefore, this is
the regime of high temperature and small chemical potential. However,
as we will see in the next section, we do not have to rely on the
matching technique if $T\gg 1$ and $T\gg\mu_1$, which encompasses the
above high temperature regime.

\subsection{Perturbative approach for a second high temperature regime}

We will now consider the linearized $\varphi_1$ equation
(\ref{eq:z1eqn}) for the case of $x_h\gg q_1$ and $x_h\gg 1$, where we
have taken $g=1$ for simplicity. As we will see, this is a second high temperature regime. 
Note that $\varphi_1$ can be expanded for small $x_h^{-1}$ as
%%%%%
\be
\varphi_1 = u_0 \Big(1 + x_h^{-1}\, u_1 + {\cal O}
(x_h^{-2})\Big)
\,.\label{expandphi}
\ee
%%%%
The function $u_0$ satisfies the following linear differential
equation
%%%%%
\be
((z^2-1)z\, u_0')' + z\, u_0=0\,,
\ee
%%%%
which has the general solution
%%%%
\be
u_0=\fft{2c_1}{\pi}\,K(1-z^2) + \fft{2c_2}{\pi}\,K(z^2)
\,.
\ee
%%%%
where $K(x)$ denotes the complete elliptic integral of the first kind.
For the solution to be regular at the horizon $z=1$, it is necessary
to set $c_2=0$. Furthermore, without loss of generality we can set the
scaling factor $c_1=1$, so that near the horizon $u_0$ behaves like
%%%%
\be
u_0=1 - \ft12 (z-1) + \ft{5}{16} (z-1)^2 + {\cal O}((z-1)^3)\,.
\ee
%%%%
Asymptotically we find that
%%%%
\be
u_0\rightarrow \fft{2[\log z-\gamma-\psi(1/2)]}{\pi z}\,,
\ee
%%%%
where $\gamma$ is Euler's constant and $\psi$ is the digamma function. Approximate 
numerical values are $\gamma\approx .58$ and $\psi(1/2)\approx -1.96$.

We can now consider the $x_h^{-1}$ corrections in $\varphi_1$. The
function $u_1$ can be expressed in terms of a functional integral as
%%%%
\bea
u_1 &=& \int_1^z \fft{v(y)}{y(y^2-1) u_0^2(y)}\, dy\,,\nn\\
v &=& I_+(z)+q_1 I_-(z)\,,\qquad I_{\pm}(z)\equiv \int_1^z
\fft{1}{y+1}\Big( y(y-1)u_0(y)u_0'(y)\pm y\,u_0^2(y)\Big) dy\,.\label{u1v}
\eea
%%%%
Setting the lower bound of the $v$ integral to unity ensures that the
solution remains regular at the horizon. Moreover, the chosen lower
bound of the $u_1$ integral guarantees that $u_1$ vanishes at $z=1$,
so that the boundary condition on $\varphi_1$ remains unchanged.

To compute the asymptotic behavior in $u_1$, evaluate
$I_{\pm}(\infty)$ in order to take the large $z$ expansion of the
$u_1$ integral. The result gives
%%%
\be
\varphi_1\rightarrow u_0-\fft{\pi (I_+(\infty)+q_1 I_-(\infty))}{2x_h z}\,,
\ee
%%%%
for large $z$. The approximate numerical values are 
$I_+(\infty)\approx 1.03$ and $I_-(\infty)\approx -3.51$. In terms of 
the coordinate $r$, the asymptotic behavior is roughly given by 
%%%%
\be
\varphi_1\rightarrow \fft{4x_h\log r}{\pi r^2}+
\fft{c}{r^2}\,,\qquad c\equiv -\fft{2x_h}{\pi}
[\log x_h+\gamma+\psi(1/2)]-\fft{\pi}{2}
[I_+(\infty)+q_1 I_-(\infty)]\,.\label{varphi1}
\ee
%%%%
Once again, the $1/r^2$ term corresponds to giving a vev to the chiral
primary operator $\Tr\,Z^2$. We normalize $\varphi_1$ by equating this
vev to the value $v_1=\sqrt{q_1(q_1+2)}$ for the case of the BPS AdS bubble, as we did in the
previous section. This enables us to find the thermal mass of
$\Tr\,Z^2$ to be
%%%%
\be
{\bar m}^2\approx \fft{\sqrt{q_1(q_1+2)}}{\log x_h}\,,\label{m}
\ee
%%%
in the limit $x_h\gg q_1$ and $x_h\gg 1$. We would like to express
this in terms of the temperature, chemical potential and $R$-charge of
the field theory. From (\ref{temp}) and (\ref{chempot}) and
(\ref{Qi}), we can express the temperature, chemical potential and
$R$-charge $Q_1$ in terms of the black hole parameters $x_h$ and $q_1$
in the regime $x_h\gg q_1$ and $x_h\gg 1$ as
%%%%
\be
T\approx \fft{\sqrt{x_h}}{\pi}\,,\qquad \mu_1\approx
\sqrt{q_1}\,,\qquad Q_1\approx \sqrt{q_1} x_h\,.
\ee
%%%%
In the grand canonical ensemble, this regime corresponds to $T\gg \mu_1$ and $T\gg 1$
and the thermal mass can be expressed as
%%%%
\be
{\bar m}^2\approx \fft{\sqrt{\mu_1^2 (\mu_1^2+2)}}{2\log T}\,.
\ee
%%%
Thus, the thermal mass increases with chemical potential and
decreases with temperature.

In the canonical ensemble, this regime corresponds to $T^3\gg Q_1$ and $T\gg 1$ and
%%%%
\be
{\bar m}^2\approx \fft{Q_1\sqrt{Q_1^2+2\pi^4 T^4}}{2\pi^4 T^4 \log T}\,.
\ee
%%%%
The thermal mass now increases with $R$-charge $Q_1$ and still
decreases with temperature.  Note that, in this regime, the results
for the thermal mass are identical for the case of three equal
charges.

Recall that there was only a small window for which the regime
discussed in the previous section satisfied the local thermodynamic
stability constraints and entropic considerations. On the other hand,
for the regime $x_h\gg q_1$ and $x_h\gg 1$, all of these constraints are
satisfied. This provides a large range of temperatures for which this
system can be reliably discussed.  Of course, backreaction and
non-linear effects would also have to be taken into account, should we
desire a more detailed treatment of the thermodynamics.

%%%%%%%%%%%%%%%%%%%%%%%%%%%%%%%%%%%%%%%%
\section{Taking backreaction into account}

In the previous section, we have explored the linearized equations of
motion for the deformation scalars $\varphi_i$.  Going beyond linear
order, these scalars will backreact on the metric through the Einstein
equation, as well as on the other matter fields through the couplings
implicit in the Lagrangian (\ref{eq:lagnew}).  In particular, the
field $\varphi_i$ acts as a source for the $A^i$ gauge fields through
the $\sinh^2\varphi_i *A^i\wedge A^i$ couplings, and the $X_i$ (or
equivalently $\phi_1$ and $\phi_2$) scalars through the modified
scalar potential (\ref{eq:spotnew}).

While the generalized backreaction equations are straightforward to
obtain, we restrict the analysis to the non-extremal generalization of
the single-charge AdS bubble.  In this case, the natural way to
parameterize the metric backreaction is to start from the (one-charge)
black hole solution (\ref{eq:blackh}) and to write
\begin{equation}
ds^2=-H^{-2/3}fdt^2+H^{1/3}(f^{-1}dr^2+r^2d\Omega_3^2),
\label{eq:brmetform}
\end{equation}
where the metric functions are, to second order in the linearization
parameter $\epsilon$,
\begin{eqnarray}
H&=&1+\fft{q_1}x+\epsilon^2h_2(x)+\cdots,\nonumber\\
f&=&1-\fft{m}{x}+g^2xH+\epsilon^2f_2(x)+\cdots
=1+g^2q_1-\fft{m}x+g^2x+\epsilon^2(f_2+g^2 x h_2)+\cdots.
\label{eq:hfexpand}
\end{eqnarray}
Here we recall that $x=r^2$ and $q_1=m \sinh^2\beta_1$, as indicated
in (\ref{eq:neharm}).  Note that we have continued to write the function
$f$ in the natural combination of $f=1-m/x+g^2xH$ plus corrections.

We now turn to the matter fields.  For the electric potential, we follow
(\ref{eq:shifta1}) and write
\begin{eqnarray}
A_t^1&=&(b^1-\coth\beta_1H^{-1})+\epsilon^2a_2(x)+\cdots\nonumber\\
&=&\coth\beta_1\left(\fft{x_h}{x_h+q_1}-\fft{x}
{x+q_1+x\epsilon^2h_2}\right)+\epsilon^2a_2+\cdots\nonumber\\
&=&-q_1\coth\beta_1\fft{x-x_h}{(x_h+q_1)(x+q_1)}+\epsilon^2
\left(\coth\beta_1\fft{x^2}{(x+q_1)^2}h_2+a_2\right)+\cdots,
\label{eq:at1corr}
\end{eqnarray}
where
$\coth\beta_1=\sqrt{1+m/q_1}=\sqrt{(g^2/q_1)(x_h+1/g^2)(x_h+q_1)}$.
Finally, for the backreaction of $\varphi_1$ on the $X_i$ scalars, we
note that it is consistent to set $X_2=X_3=1/\sqrt{X_1}$, since
$\varphi_1$ sources $X_2$ and $X_3$ in an identical manner.  We then
take a multiplicative parameterization of $X_1$:
\begin{equation}
X_1=H^{-2/3}(1+\epsilon^2\chi_2(x))=\left(1+\fft{q_1}x\right)^{-2/3}
\left(1+\epsilon^2\left(\chi_2-\ft23\fft{x}{x+q_1}h_2\right)\right)+\cdots.
\end{equation}

For $\varphi_1$ of order $\epsilon$, its backreaction on the metric
fields $h_2$ and $f_2$, the electric potential $a_2$ and scalar
$\chi_2$ is then governed by the set of inhomogeneous second-order
equations
\begin{eqnarray}
\label{eq:hbr}
&&[x^2h_2'+2q_1\chi_2]'=-x(x+q_1)\varphi_1'^2
-\fft{q_1x(x_h+1/g^2)}{(x_h+q_1)(x+x_h+1/g^2+q_1)^2}\varphi_1^2,\\
\label{eq:fbr}
&&(x^2f_2)''+g^2[x^3h_2'+2x^2h_2]'+g^2[x^2h_2'+2q_1\chi_2]\nonumber\\
&&\kern12em
=\fft{g^2x(x+x_h+1/g^2+2q_1+q_1(x_h+1/g^2)/(x_h+q_1))}{x+x_h+1/g^2+q_1}
\varphi_1^2,\\
\label{eq:abr}
&&((x+q_1)^2a_2')'=\coth\beta_1x(x+q_1)\left(\varphi_1'^2-\fft{q_1}{(x_h+q_1)
(x+x_h+1/g^2+q_1)^2}\varphi_1^2\right),\\
\label{eq:xbr}
&&[(x(x-x_h)(x+x_h+1/g^2+q_1)\chi_2')'+x\chi_2]
+[2xf_2/g^2+x^2h_2]'+[x^2h_2'+2q_1\chi_2]=x\varphi_1^2.\qquad
\end{eqnarray}
In addition, the constraint equation coming from the Einstein equations
gives rise to a rather cumbersome first-order equation
\begin{eqnarray}
&&\fft{2g^2q_1}{x+q_1}(x-x_h)(x+x_h+1/g^2+q_1)\chi_2'+2g^2q_1
\left(\fft{(x_h+1/g^2)(x_h+q_1)}{(x+q_1)^2}-1\right)\chi_2\nonumber\\
&&-\fft{x(3x+2q_1)}{x+q_1}f_2'+\left(\fft{q_1^2}{(x+q_1)^2}-3\right)f_2
+2q_1\coth\beta_1a_2'\nonumber\\
&&-\left(4g^2x^2-\fft{xx_h(1+g^2(x_h+q_1))}{x+q_1}\right)h_2'
-\left(4g^2x-\fft{q_1x_h(1+g^2(x_h+q_1))}{(x+q_1)^2}\right)h_2\nonumber\\
&&\kern4em=-g^2x(x-x_h)(x+x_h+1/g^2+q_1)\varphi_1'^2\nonumber\\
&&\kern5em-\fft{g^2x(x+x_h+1/g^2+2q_1+q_1(x_h+1/g^2)/(x_h+q_1))}
{(x+x_h+1/g^2+q_1)}\varphi_1^2.
\label{eq:firstoe}
\end{eqnarray}

We observe that the metric backreaction equations (\ref{eq:hbr}) and
(\ref{eq:fbr}), along with the $X_1$ equation (\ref{eq:xbr}), are
coupled in a non-trivial manner.  On the other hand, the electric
potential equation (\ref{eq:abr}) is independent, and can be solved by
quadratures:
\bea
a_2(x)&=&\coth\beta_1\int^x \fft{dx'}{(x'+q_1)^2}\int^{x'} dx''\, x''(x''+q_1)
\times\nn\\
&&
\qquad\qquad\qquad\qquad
\left(\varphi_1'^2(x'')-\fft{q_1}{(x_h+q_1)(x''+x_h+1/g^2+q_1)^2}
\varphi_1^2(x'')\right).
\eea
Note that the indefinite integrals allow for the addition of an
arbitrary homogeneous solution
\begin{equation}
a_2=k_1+k_2\fft1{x+q_1}+\cdots.
\end{equation}
These constants may be fixed by demanding that the leading asymptotic
behavior of the solution be unchanged by the backreaction.  In
particular, for the electric potential we demand the vanishing of any
$1/x$ correction to $A_t^1$ (which would modify the $R$ charge).  We
also insist that the potential continues to vanish at the horizon,
$A_t^1(x_h)=0$, so that $A^1$ remains normalizable at the horizon.
Note, however, that these two conditions do not fix $k_1$ and $k_2$
directly, since $h_2$ also enters into the correction to the
potential, as indicated in (\ref{eq:at1corr}).  Assuming (as we
demonstrate below) that $h_2$ falls off faster than $1/x$ at infinity,
the above requirements then lead to
\bea
a_2(x)&=&-\coth\beta_1\biggl[
\int_{x_h}^x\fft{dx'}{(x'+q_1)^2}\int_{x'}^\infty dx''\,x''(x''+q_1)
\times\nn\\
&&
  \left(\varphi_1'^2(x'')-\fft{q_1}{(x_h+q_1)(x''+x_h+1/g^2+q_1)^2}
\varphi_1^2(x'')\right)
-\fft{x_h^2}{(x_h+q)^2}h_2(x_h)\biggr].
\eea
From the asymptotic behavior of $\varphi_1$ as given in (\ref{eq:vphias})
\begin{equation}
\varphi_1\sim\fft{c_1+c_2\log x}x+\cdots,\qquad x\to\infty,
\label{eq:vphias2}
\end{equation}
we obtain
\begin{eqnarray}
a_2&\sim&a_2(\infty)+\fft{\coth\beta_1}{x^2}\biggl[-c_2(c_1+c_2)
+\fft{x_h}{x_h+q_1}\left(\ft12c_1^2+\ft32c_1c_2+\ft74c_2^2\right)
\nonumber\\
&&\kern6em+\left(-c_2^2+\fft{x_h}{x_h+q_1}c_2\left(c_1+
\ft32c_2\right)\right)
\log x+\ft12\fft{x_h}{x_h+q_1}c_2^2\log^2x+\cdots\biggr].\qquad
\label{eq:a2bra}
\end{eqnarray}
The constant $a_2(\infty)$ contributes to a shift in the value of the
electric potential (or, equivalently, the $R$ charge chemical
potential) compared to the zeroth-order solution.

Turning to the other fields $\{h_2,f_2,\chi_2\}$, the coupled system
of backreaction equations (\ref{eq:hbr}), (\ref{eq:fbr}) and
(\ref{eq:xbr}) do not appear to admit a straightforward solution.
Nevertheless, some general properties are evident.  Firstly, in the
limit of vanishing electric charge, $q_1\to0$, the metric equations
(\ref{eq:hbr}) and (\ref{eq:fbr}) become self-contained.  In this
limit, (\ref{eq:hbr}) may be integrated twice to obtain $h_2$ and, in
turn, knowledge of $h_2$ allows (\ref{eq:fbr}) to be integrated for
$f_2$.  Secondly, even with $q_1\ne0$, the combination of the metric
equations admit first integrals, thus reducing the order of the
coupled system.  In fact, using the first-order equation,
(\ref{eq:firstoe}), along with the backreaction equations for
$\{h_2,f_2,a_2\}$, we may obtain the homogeneous equation
\begin{eqnarray}
&&g^2x_h(x_h+1/g^2+q_1)\left(\fft{x}{x+q_1}h_2\right)'+g^2x_h(x^2h_2')'
+\left(x^2f_2'-\fft{x^2}{x+q_1}f_2\right)'\nonumber\\
&&\quad+q_1\coth\beta_1\left(a_2+\fft{(x+q_1)(x-x_h)}{x_h+q_1}a_2'\right)'
+2g^2q_1(x_h+1/g^2)\left(\fft{x-x_h}{x+q_1}\chi_2\right)'=0,
\end{eqnarray}
which yields yet another constant of integration (which is presumably
related to the conserved energy of the static
gravitational system).  Finally, we note from (\ref{eq:xbr}) that the
$\chi_2$ scalar fluctuations satisfy an equation of motion which is of
similar form to the $\varphi_i$ equation (\ref{eq:seom}).  This should
not be of much surprise, as small scalar fluctuations are naturally
governed by the Klein-Gordon equation, which in the present background
takes on the form
\begin{equation}
(x^2f\Phi')'-\ft14M^2xH^{1/3}\Phi=0,
\end{equation}
for a scalar $\Phi$.  (Note that, in a background with non-zero $R$
charge, the supergravity potential (\ref{eq:spotnew}) leads to
position-dependent `masses' for both the $\varphi_i$ and $X_i$
scalars.)  Since $\varphi_i$ and $X_i$ both originate from the
$T_{ij}$ tensor of the sphere reduction \cite{Cvetic:2000nc}, they all
have $E_0=2$, and hence share a common value of mass $M^2=-4g^2$ (at
least when the charges are turned off).  Regardless of the details, we
expect that the $\chi_2$ fluctuations are likewise described by a Heun
equation.  Hence similar difficulties to those we encountered in the
previous section arise when obtaining explicit solutions for the
backreaction of $\varphi_1$ on $\chi_2$.

Because the system of backreaction equations is linear, the
inhomogeneous solution can in principle be obtained by variation of
parameters, so long as the fundamental matrix ({\it i.e.}~the complete
set of linearly independent solutions to the homogeneous system) is
known.  More precisely, we may rewrite the second-order equations in
coupled first-order form.  Then the solution to the system of $n$
first-order equations
\begin{equation}
\vec f\,'(x)=\vec A(x)x+\vec B(x),
\end{equation}
may be formally expressed as
\begin{equation}
\vec f(x)=\Phi(x)\vec f_0+\Phi(x)\int_{x_0}^x\Phi^{-1}(x')\vec B(x')\,dx',
\end{equation}
where $\Phi(x)$ is the fundamental matrix satisfying $\Phi'(x)=\vec
A(x)\Phi(x)$ and normalized according to $\Phi(x_0)=I$.  This formal
solution is not particularly useful, since in the present case
$\Phi(x)$ would be a $7\times7$ matrix, corresponding to the freedom
of four second-order equations (\ref{eq:hbr}) through (\ref{eq:xbr})
minus the first-order constraint (\ref{eq:firstoe}).  Nevertheless, we
note that five out of the seven linearly independent solutions to the
homogeneous system are easy to obtain:
\begin{equation}
\begin{tabular}{l|llll}
&$h_2$&$f_2$&$a_2$&$\chi_2$\\
\hline
1:&&&$1$\\[4pt]
2:&$\displaystyle-
\fft{2q_1\coth\beta_1}{m x}\kern2em$&&$\displaystyle\fft1{x+q_1}$
\kern2em&\\[12pt]
3:&$\displaystyle\fft{q_1}x$&$\displaystyle-\fft{m}{x}$\\[12pt]
4:&$\displaystyle\fft{q_1}{x^2}$&$\displaystyle-\fft{m}{2x^2}
-\fft{3m+4q_1}{2q_1x}$&&$\displaystyle\fft1x$\\[12pt]
5:&$\displaystyle\fft32$&$\displaystyle-g^2(x+q_1)+\fft{7m+4q_1}{2x}
\kern2em$&&$1$
\end{tabular}
\label{eq:homos}
\end{equation}
where $m=x_h(1+g^2(x_h+q))$ as obtained from (\ref{eq:muexp}).  The
remaining two solutions appear to involve non-trivial Heun functions
pertaining to the $\chi_2$ system.

Our main interest in studying the backreaction is to obtain the
asymptotic form of the corrections so we may discern what effects
turning on $\varphi_1$ may have on the conserved quantities such as
mass and $R$ charge.  For the latter, we have already seen from
(\ref{eq:a2bra}) that we may hold the charge $q_1$ fixed, even as we
turn on $\varphi_1$.  For the mass, we turn to an asymptotic expansion
of the coupled system (\ref{eq:hbr}), (\ref{eq:fbr}) and
(\ref{eq:xbr}) as $x\to\infty$.  From the asymptotic $\varphi_1$
behavior (\ref{eq:vphias2}), we obtain
\begin{eqnarray}
h_2&\sim&\fft1{x^2}\biggl[
-\ft12c_1^2-\ft12c_1c_2-\ft34c_2^2+q\left(\chi_1+\ft12\chi_{11}\right)
\nonumber\\
&&\kern4em+\left(-c_2\left(c_1+\ft12c_2\right)+q_1\chi_{11}\right)\log x
-\ft12c_2^2\log^2x\biggr]+\cdots,\nonumber\\
f_2&\sim&\fft1{x^2}\biggl[\ft23q_1\left(\chi_1+\ft13\chi_{11}\right)
+\ft13c_2\left(2c_1+\ft53c_2\right)-\fft{x_h(1-g^2q_1)}{x_h+q_1}
\left(\ft13c_1^2+\ft89c_1c_2+\ft{26}{27}c_2^2\right)\nonumber\\
&&\kern4em+\left(\ft23q_1\chi_{11}+\ft23c_2^2
-\ft29\fft{x_h(1-g^2q_1)}{x_h+q_1}c_2(3c_1+4c_2)\right)\log x\nonumber\\
&&\kern4em-\ft13\fft{x_h(1-g^2q_1)}{x_h+q_1}c_2^2\log^2x
\biggr]+\cdots,\nonumber\\
\chi_2&\sim&\fft1x\left(\chi_1+\chi_{11}\log x\right)
+\fft1{x^2}\left[c_2\left(c_1+\ft32c_2\right)
+\fft{\chi_{11}}{g^2}+c_2^2\log x\right]+\cdots.
\label{eq:brsys}
\end{eqnarray}
Note that here we have used the freedom expressed in (\ref{eq:homos})
to set the leading order terms in $h_2$ and $f_2$ to zero.  The
content of the residual two homogeneous solutions are incorporated
through the constants $\chi_1$ and $\chi_{11}$, which are related to the shift in
the asymptotic profile of the scalar $X_1$ at infinity.  Of course,
the contribution for the full set of homogeneous solutions may have to
be added back in to satisfy the desired boundary conditions.  For the
gauge potential, we have argued that it is natural to demand
$A_t^1(x_h)=0$.  Likewise, here it would be appropriate to set
$f_2(x_h)+g^2 x_h h_2(x_h)=0$ so that $f(x)$ as given in
(\ref{eq:hfexpand}) continues to vanish at $x_h$ when the backreaction
is included.

Although we have explored the backreaction of $\varphi_1$ on the other
fields of the system, we can also see from (\ref{eq:brsys}) that the
$X_1$ scalar may be deformed as well, through $\chi_1$ and
$\chi_{11}$.  Presumably, this would allow a wider class of thermal
black hole solutions with scalar hair.  However, assuming our goal is
to thermalize the BPS bubble, one may presumably set
$\chi_1=\chi_{11}=0$ directly without any major concern.  At the same
time, we note an interesting feature of the non-extremal bubble
solution.  While the $R$-charged black hole (\ref{eq:blackh}) and the
BPS bubble (\ref{eq:adsbubble}) both have the $X_1$ scalar satisfying
$X_1=H_1^{-2/3}$ (for the one-charge case), this condition can no
longer be maintained when the bubble is thermalized, as evidenced by
the non-vanishing backreaction on $\chi_2$.

We also note that, according to (\ref{eq:brsys}), it appears that the
backreaction can be adjusted so that it has no effect on the mass of
the black hole. This is because, at least heuristically, the mass can
be read off from the $1/r^2\equiv1/x$ terms in the metric functions
$H_1$ and $f$ (since we are working in five dimensions), and both such
terms are absent in (\ref{eq:brsys}).  It may turn out, however, that
boundary conditions at the horizon would feed in some of the
homogeneous solution of the third type in (\ref{eq:homos}), thus
leading to a possible shift in the mass.  Of course, such issues
cannot be properly examined in the linearized analysis, and would have
to await a full (possibly numerical) solution to be addressed.
Nevertheless, the heuristic concept of mass can be made rigorous, and
this is what we turn to next.

\section{Mass of the non-BPS bubbles}

   The definition, and calculation, of the mass of an asymptotically
AdS spacetime is somewhat more subtle than it is in an asymptotically
flat spacetime.  One can no longer use the ADM definition, which
assumes an asymptotically Minkowski region \cite{adm}.  A
generalisation of the ADM procedure, in which one decomposes the
metric as the sum of an AdS background plus deviations, was introduced
in \cite{abodes} by Abbott and Deser.  Effectively, one is making an
infinite background AdS subtraction from a divergent boundary
integral.  The presence of scalar fields in the solution, such as one
has in the supergravity black holes and bubbles, can complicate the
application of this AD procedure considerably, because of the inherent
ambiguities in the separation of the metric into background plus
deviations.  Some discussion of the AD approach, and calculations for
higher-dimensional rotating black holes, can be found in
\cite{dekate}.

   A procedure for calculating the mass of asymptotically AdS
spacetimes that avoids all the problems inherent in making a split
into background and deviation was introduced by Ashtekar, Magnon and
Das \cite{ashmag,ashdas}.  This is based on a conformal definition
that expresses the mass in terms of an integral of certain components
of the Weyl tensor over the spatial conformal boundary at infinity.
Since the metric approaches AdS asymptotically, the integrand falls
off and the integral is inherently well-defined.  This AMD method was
applied in \cite{gibperpop} to the calculation of the masses of
higher-dimensional rotating AdS black holes in general relativity, and
in \cite{chenlupope} this was extended to the case of rotating black
holes in gauged supergravities.

Alternatively, the boundary counterterm method may be used to
calculate the mass of configurations in an AdS background
\cite{Henningson:1998gx,Balasubramanian:1999re,Emparan:1999pm,%
Kraus:1999di,deBoer:1999xf}.  This notion of holographic
renormalization is particularly natural in the context of AdS/CFT,
where the addition of boundary counterterms in AdS has a natural
counterpart in the addition of renormalization counterterms in the
dual field theory.  Furthermore, the boundary counterterm method has
the advantage that it regulates divergences not just in the mass, but
also in the on-shell gravitational action which is dual to the
thermodynamic potential of the CFT.  The boundary counterterm method
was used in \cite{Buchel:2003re,Liu:2004it,Batrachenko:2004fd} to
investigate the mass of the (undeformed) $R$-charged black holes
(\ref{eq:blackh}).

   Here, however, we shall use the AMD method to discuss the masses of
BPS bubble metrics and their non-extremal deformations. Note that, in
this definition of mass, pure AdS has by construction zero energy.
This is in contrast with holographic renormalization, which naturally
assigns non-zero energy to the vacuum (which is viewed as dual to the
Casimir energy of the CFT on $S^3\times\mathbb R$).  We begin by
briefly summarising the AMD procedure, drawing on material presented
in \cite{gibperpop,chenlupope}.

    Consider an asymptotically AdS bulk spacetime $\{X,g\}$ of
dimension $D$, equipped with a conformal boundary $\{\del X, \bar
h\}$.  It admits a conformal compactification $\{\bar X, \bar g\}$ if
$\bar X =\sqcup \del X$ is the closure of $X$, and the metric $\bar g$
extends smoothly onto $\bar X$ where $\bar g = \Omega^2\, g$ for some
function $\Omega$ with $\Omega>0$ in $X$ and $\Omega=0$ on $\del X$,
with $d\Omega\ne0$ on $\del X$.  One might, for example, take
%%%%%
\be
\Omega= \fft{l}{y}\,,\label{omegay}
\ee
%%%%%
where in the asymptotic region the metric approaches AdS with
$R_{\mu\nu}=l^{-2}\, g_{\mu\nu}$.  (For the solutions in gauged
supergravity that we shall consider, $l=1/g$, where $g$ here denotes
the gauge-coupling constant.)  Since $\Omega$ is determined only up to
a factor, $\Omega\rightarrow f\, \Omega$, where the function $f$ is
non-zero on $\del X$, the metric $\bar g$ on $\bar X$ and its
restriction $\bar h=\bar g|_{\del X}$ are defined only up to a
non-singular conformal factor.  The conformal equivalence class
$\{\del \bar X,\bar h\}$ is called the conformal boundary of $X$.  If
$\bar C^\mu{}_{\nu\rho\sigma}$ is the Weyl tensor of the conformally
rescaled metric $\bar g_{\mu\nu}= \Omega^2\, g_{\mu\nu}$, and $\bar
n_\mu \equiv \del_\mu \Omega$, then in $D$ dimensions one defines
%%%%
\be
\bar{\cal E}^\mu{}_{\nu}= l^2 \Omega^{D-3}\, \bar n^\rho \, \bar n^\sigma\,
    \bar C^\mu{}_{\rho\nu\sigma}\,.\label{cale}
\ee
%%%%%
This is the electric part of the Weyl tensor on the conformal
boundary.  The conserved charge $Q[K]$ associated to the asymptotic
Killing vector $K$ is then given by
%%%%%
\be
Q[K]= \fft{l}{8\pi\, (D-3)}\, \oint_{\Sigma}\,
       \bar{\cal E}^\mu{}_{\nu}\, K^\nu\, d\bar\Sigma_\mu\,,\label{conffor}
\ee
%%%%%%
where $d\bar\Sigma_\mu$ is the area element of the $(D-2)$-sphere
section of the conformal boundary.  (The derivation of (\ref{conffor})
is discussed in \cite{ashmag,ashdas}).  Note that the expression
(\ref{conffor}) is invariant under the non-singular conformal
transformations of the boundary metric that we discuss above.  Thus,
one may take for $\Omega$ any conformal factor that is related to
(\ref{omegay}) by a non-singular multiplicative factor.

   In order to define the energy, one takes $K=\del/\del t$, where $t$
is the time coordinate appearing in the asymptotically AdS form
%%%%%
\be
ds^2 = -(1+y^2\, l^{-2})\, dt^2 + \fft{dy^2}{1+ y^2\, l^{-2}} +
    y^2 \, d\Omega_{D-2}^2\label{asads}
\ee
%%%%%
of the metric under investigation.  The energy (or mass) is then given
by
%%%%%
\be
E = \fft{l}{8\pi\, (D-3)}\, \oint_{\Sigma} \bar{\cal E}^t{}_t\,
    d\bar\Sigma_t\,.\label{mass}
\ee
%%%%%

    For our present discussion, we need to apply (\ref{mass}) to the
class of five-dimensional metrics given by
%%%%%
\be
ds_5^2 = - {\cal H}^{-2/3}\, f dt^2 + 
{\cal H}^{1/3}(f^{-1}\, dr^2 + r^2 d\Omega_3^2)\,,
\label{massmet}
\ee
%%%%%
where ${\cal H}$ and $f$ are functions only of $r$, and $d\Omega_3^2$
is the metric on the unit 3-sphere.  It is convenient, as usual, to
define $x=r^2$.  From (\ref{mass}) we then find that the mass is given
by
%%%%%
\be
E=\lim_{x\to\infty} \fft{\pi g^2 x^2}{16 f {\cal H}^2}\, 
\Big[(1-f+ x f'-2x^2 f'') {\cal H}^2 +
  x^2(3f'{\cal H}{\cal H}'- 3f {{\cal H}'}^2 
+ 2f {\cal H} {\cal H}'')\Big]\,,\label{massform}
\ee
%%%%%
where primes denote derivatives with respect to $x$, and we have
taken the five dimensional Newton's constant $G_5=1$.

   Before applying this mass formula to the non-BPS bubbles that we
have been investigating in this paper, it is instructive to consider
some BPS bubble examples.  For the spherically-symmetric 1-charge bubble
constructed in \cite{Lin:2004nb}, the metric functions in (\ref{massmet})
are given by
%%%%%
\be
{\cal H} =H_1\,,\qquad f= 1+ g^2 x H_1\,,\qquad 
H_1= \fft{\sqrt{g^4 x^2 + 2(g^2 q_1+1)g^2 x + 1} -1}{g^2 x}\,,
\ee
%%%%%
where $q_1$ is the electric charge parameter.  Substituting into 
(\ref{massform}), we find that the mass is 
given by
%%%%%
\be
E= \fft{\pi\, q_1}{4}\,.
\ee
%%%%%

   For the more general case of 3-charge BPS bubbles, which were 
constructed in \cite{Chong:2004ce}, we have
%%%%%
\be
{\cal H}=H_1 H_2 H_3\,,\qquad f= 1 + g^2 x H_1 H_2 H_3\,.
\ee
%%%%%
In this case the explicit solution for the three functions $H_i$ is
not known, but at large $x$ they take the form
%%%%%
\be
H_1= 1+ \fft{q_i}{x} + \cdots\,.
\ee
%%%%%
Substituting into (\ref{massform}), we find that the mass is given by
%%%%%
\be
E = \fft{\pi(q_1+q_2+q_3)}{4}\,.
\label{eq:bps3mass}
\ee
%%%%% 

   As a further example, one finds from (\ref{massform}) that
the 3-charge non-extremal black holes (\ref{eq:blackh}) in 
five-dimensional gauged supergravity have mass given by
%%%%%
\be
E=\fft{\pi m}{8 }\sum_i \cosh2\beta_i
=\fft\pi4\left(\fft32m+\sum_iq_i\right)\,.
\ee
%%%%%
In the limit when $m$ is taken to zero, this black hole result reduces
to that of the BPS bubble, (\ref{eq:bps3mass}).

In general, for a gravitational background parameterized by the
asymptotic behavior
\begin{eqnarray}
\mathcal H&\sim&1+\fft{\overline h_1}x+\fft{\overline h_2}{2x^2}
+\cdots,\nonumber\\
f&\sim&1+g^2x\mathcal H+\fft{\overline f_1}x+\fft{\overline f_2}{2x^2}
+\cdots,\label{eq:calhfasy}
\end{eqnarray}
we find that application of the mass formula (\ref{massform}) gives
simply
\begin{equation}
E=\fft\pi8\left(2\overline h_1-3\overline f_1\right).
\label{eq:eh1f1}
\end{equation}
This demonstrates that the mass indeed receives contributions only
from the $1/x$ terms in the expansion of $\mathcal H$ and $f$, as
alluded to at the end of the previous section.  In reality, the
expansion in (\ref{eq:calhfasy}) may also include log terms.  However,
so long as the logs are confined to the $1/x^2$ and higher terms, the
mass remains finite and unchanged from (\ref{eq:eh1f1}).

For the linearized bubbling AdS black hole solution, the metric
functions $\mathcal H\equiv H_1$ and $f$ are given by the backreaction
expansions (\ref{eq:hfexpand}).  As a result, we obtain for the mass
\begin{equation}
E=\fft\pi4\left[\fft32m+q_1+\lim_{x\to\infty}x\left(h_2(x)-
\fft32f_2(x)\right)\right].
\end{equation}
Note that we have already assumed that the backreaction functions
$h_2$ and $f_2$ vanish at infinity
\begin{equation}
h_2(\infty)=f_2(\infty)=0.
\end{equation}
Based on the backreaction expansion (\ref{eq:brsys}), and taking into
account the possible contribution of the homogeneous solutions in
(\ref{eq:homos}), we see that the masses of the bubbling AdS black
holes remain finite, at least for arbitrary linearized deformations.
Unfortunately, however, the local analysis leading to (\ref{eq:brsys})
is insufficient in itself to determine how the mass varies as the
$\varphi_1$ deformation is turned on while keeping, say, temperature
and $R$-charge fixed.

%%%%%%%%%%%%%%%%%%%%%%%%%%%%%%%%%
\section{A numerical approach}

As in much of the rest of this paper, our interest is in non-BPS
bubbles.  However, we also numerically explore some features of the
BPS bubbles of \cite{Chong:2004ce}.

\subsection{Non-BPS bubbles}

         There are two types of AdS bubbles.  The first type can be
referred to as solitonic AdS bubbles, of which the BPS bubbles are a
subset.  The corresponding geometry is completely regular and
horizon-free. It interpolates between AdS spacetime at asymptotic
infinity and Minkowskian spacetime at short distance. The second type
is the thermal AdS bubble, for which the geometry contains a singular
point surrounded by a horizon.  In the above sections, we have mainly
looked at the linearized solution of a subset of thermal AdS bubbles,
which we have referred to as bubbling AdS black holes.  We can
demonstrate the existence of both types of AdS bubbles through
numerical analysis.

\bigskip\bigskip
{\noindent \underline{Solitonic AdS bubbles}}
\bigskip

Solitonic AdS bubbles have no horizon, and may be described by a
global radial coordinate $x\in[0,\infty)$.  Focusing on the one-charge
spherically symmetric system, the Lagrangian (\ref{eq:lagnew}) gives
rise to a coupled set of non-linear ordinary differential equations
for the metric functions $f$ and $H_1$ as well as the matter fields
$X_1$, $\varphi_1$ and $A_t^1$.  In general, the equations of motion,
including the Einstein equations, gives rise to five second-order
equations (one for each function).  However, there is also a first-order 
`energy' or constraint equation arising from the Einstein
equations.  As a result, any general solution may essentially be
specified by nine constants.  Not all such solutions are physically
independent, however.  Even with the metric given in the form
(\ref{eq:brmetform})
\begin{equation}
ds^2=-H(x)^{-2/3}f(x)dt^2+H(x)^{1/3}\left(\fft{dx^2}{4xf(x)}
+xd\Omega_3^2\right),
\end{equation}
there remains a residual coordinate transformation
\begin{equation}
t\to\lambda^{-1}t,\qquad x\to\lambda x,\qquad H\to\lambda^{-3}H,\qquad
A_t^1\to\lambda A_t^1,
\label{eq:scaling}
\end{equation}
leaving the form of the metric invariant.  This reduces the nine constants
to eight physical parameters of the solution.

Of course, most of these solutions may be singular as $x\to0^+$.  For
solitonic AdS bubbles, we necessarily demand regularity at the origin,
and hence may set up initial conditions by obtaining the Taylor
expansion of the solution near $x\rightarrow 0^+$.  For small $x$, up
to linear order, this expansion is given by
%%%%
\bea
H_1&=&H_0 -\ft12 H_0^2 A_0^2 (\cosh^2\varphi_1^0-1) x+\cdots\,,\nn\\
X_1&=&X_0 + \ft13 H_0^{\fft13} (1-\cosh\varphi_1^0 X_0^\fft32 +
          (\cosh^2\varphi_1^0-1) X_0^3)x+\cdots\,,\nn\\
\cosh\varphi_1 &=& \cosh\varphi_1^0 -\ft12 (\cosh^2\varphi_1^0-1)
H_0^{\fft13} (A_0^2 \cosh\varphi_1^0 H_0^{\fft23} +
2\sqrt{X_0} - \cosh\varphi_1^0\, X_0^2)x+\cdots\,,\nn\\
A_1&=&A_0 + \ft12 A_0(\cosh^2\varphi_1^0-1) X_0^2 H_0^\fft13\, x+
\cdots\,,\nn\\
f&=&1 +\fft{(\cosh^2\varphi_1^0 -1)A_0 H_0^2 X_0 +
H_0^\fft43 (2 + 4 X_0^\fft32 \cosh\varphi_1^0 + (\cosh^2\varphi_1^0-1)
X_0^3)}{6H_0 X_0}x+\cdots\,.\qquad
\eea
%%%%
Thus, by imposing regularity, we see that the solution ends up being
parameterized by only four constants, namely $H_0$, $A_0$, $X_0$ and
$\varphi_1^0$.  One of the parameters is trivial, owing
to the residual symmetry (\ref{eq:scaling}).
Note that $f_0\equiv f(x=0)$ is fixed to be unity by
the equations of motion, and by regularity of the spatial slice for
shrinking $S^3$.

\begin{figure}[t]
\begin{center}
\includegraphics[width=3.4truein]{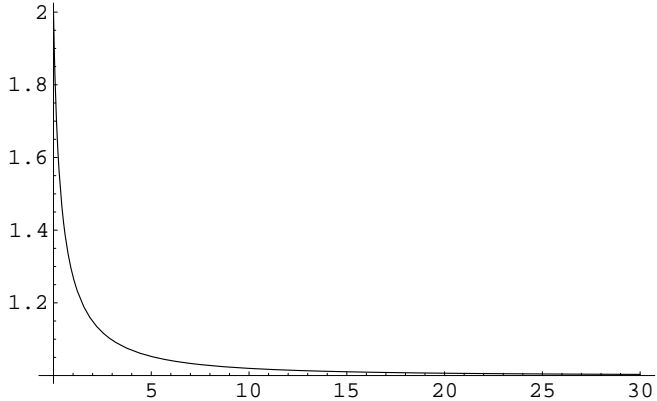}
\end{center}
\caption{{Plot of $\cosh\varphi_1$ as a function of $x$ for a non-BPS
soliton bubble with parameters given by (\ref{eq:sbnpara})}.}
\label{fig:soliton}
\end{figure}

To obtain the BPS bubble, we may impose the first-order condition
$\cosh\varphi_1=(x H_1)'$ given in (\ref{eq:adsbubble}).  This reduces
the four constants down to one according to
%%%
\be
\cosh\varphi_1^0=H_0\,,\qquad
A_0=\fft1{H_0}\,,\qquad X_0=H_0^{-\fft23}\,,
\label{eq:bpsbctrs}
\ee
%%%%
and precisely gives rise to the BPS bubbles that preserve $\ft12$ 
supersymmetry, and with charge $q=H_0-1$.  On the other hand, general
solitonic (but non-BPS) bubbles may be obtained by relaxing any
or all of the constraints in (\ref{eq:bpsbctrs}).  While we do not
explore this parameter space in detail, we can numerically demonstrate
that non-BPS solitonic bubbles can also exist. For example, taking
\begin{equation}
H_0=8,\qquad A_0=1/8,\qquad X_0=1/4,\qquad \cosh\varphi_1^0=2
\label{eq:sbnpara}
\end{equation}
gives rise to a smooth solution with $\varphi_1$ profile given in
Fig.~\ref{fig:soliton}.  Taking the scaling symmetry
(\ref{eq:scaling}) into account, these non-BPS bubbles are
parameterized by three physical parameters, presumably the $R$-charge,
as well as two scalar `charges' describing $s$-wave excitations of the
$X_1$ and $\varphi_1$ scalars in AdS.  These solutions can be viewed
as continuous non-BPS deformations of the BPS AdS bubbles.

%%%%
%%%%

\bigskip\bigskip
{\noindent \underline {Thermal AdS bubbles}}
\bigskip

We now turn to AdS bubbles with horizons.  Since the horizon appears
as a coordinate singularity, we restrict the numerical solution to
cover the outside region $x\in[x_h,\infty)$.  In this case, the
generic regular solution near the horizon may be specified by six
parameters, namely the location of the horizon $x=x_h$, the initial
values of the functions $X_1(x_h)$, $H_1(x_h)$, $\cosh\varphi_1(x_h)$
and the slopes $f_1$ and $a_1$ of $f(x)$ and $A_1(x)$, which both much
approach zero as $x\to x_h$.  Note that, taking the scaling symmetry
(\ref{eq:scaling}) into account, we may set $x_h$ to a generic value
(say $x_h=1$).  This indicates that the solutions may be specified by
five physical parameters.  However, we find it convenient to allow
$x_h$ to remain free in the numerical work.

\begin{figure}[t]
\begin{center}
\includegraphics[width=3.4truein]{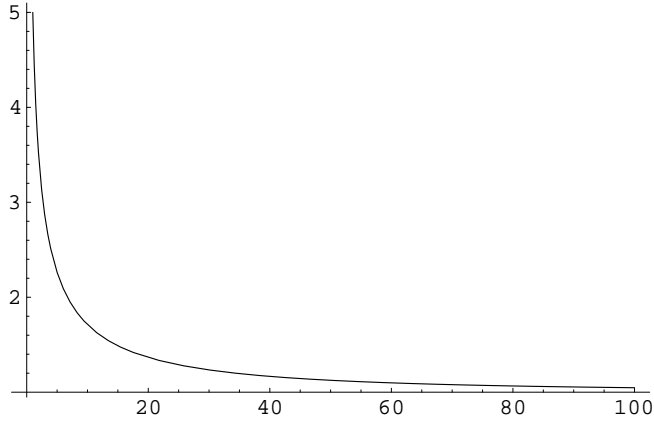}
\end{center}
\caption{{Plot of $\cosh\varphi_1$ as a function of $x$ for a thermal bubble.
We have taken $q=10$, $x_h=1$ and $\cosh\varphi_1^0=5$.}}
\label{fig:thermal}
\end{figure}

If we let $\cosh\varphi_1=1$, then we turn off the bubble parameter
and the solution reduces to that of a charged AdS black hole.  It is
worth remarking that the previously-known charged AdS black hole
solution, in which the scalar $X_1$ does not have an independent
charge parameter, is not the unique spherically symmetric black hole.
Numerical analysis shows that the generic black hole is, in fact,
characterized by three parameters: the mass, the $R$-charge and the
scalar charge of $X_1$.  The thermal AdS bubble solution, which also
turns on $\varphi_1$, is then characterized by four parameters.
(It remains unclear what the elusive fifth parameter is.
Possibly it could be removed by yet
another residual gauge transformation.)  Using a numerical approach,
we can demonstrate the existence of these solutions.  In particular,
we present an AdS bubble solution that can be viewed as a deformation
of the previously-known AdS black hole.  In this case, the initial
conditions are specified by the following:
%%%%%
\be
H_0=1 + \fft{q}{x_h}\,,\qquad X_0=H_0^{-\ft23}\,,\qquad
f_1=\fft{1+q+2x+h}{x_h}\,,\qquad a_1^2=\fft{q(x_h+1)}{(x_h+q)^3}\,.
\ee
%%%%
For $\cosh\varphi_1=1$, this leads precisely to the previously-known
charged AdS black hole.  We can turn on $\varphi_1$ and numerical
analysis indicates that the solution exists provided that
$1\le\cosh\varphi_1 < H_0$.  In Fig.~\ref{fig:thermal} we present a
plot of $\cosh\varphi_1$ for $q=10$, $x_h=1$ and $\cosh\varphi_1^0=5$.

%%%%
%%%%

\subsection{BPS bubbles}

  In \cite{Chong:2004ce}, defining equations for multi-charge BPS
bubbles were obtained in $D=4, 5, 6$ and 7 AdS gauged supergravities.
Those equations in general do not admit analytical solutions.  Here we
shall report that numerical analysis indicates that smooth bubble
solutions exist in all of these cases.  In all of these solutions, the
coordinate $x$ (which is denoted as $R$ in \cite{Chong:2004ce}) runs
from 0, where the metric is Minkowskian, to asymptotic AdS spacetime
at $x=\infty$.

     In $D=5$, the functions $H_i$ describing the general three-charge BPS 
bubble solutions satisfy the nonlinear equation (\ref{eq:susyeom}), which
may be rewritten as
\begin{equation}
\xi_i''=-g^2\fft{(\xi_i'^2-1)\xi_j\xi_k}{x^2+\xi_1\xi_2\xi_3},
\label{eq:newbubeqn}
\end{equation}
where $i\ne j\ne k$, and where we have defined $\xi_i\equiv x H_i$.
The one-charge case is given by $\xi_2=\xi_3=x$, while the two-charge
case is given by $\xi_3=x$.

Regularity of the BPS bubble demands that the $\xi_i$ vanish linearly
(or equivalently that the $H_i$ approach constants) as $x\to0$.  Taking
this into account, we may develop a Taylor expansion around $x=0$:
%%%%
\be
\xi_i\equiv xH_i=a_ix - \ft12 a_j a_k (a_i^2-1)\, x^2
 + \ft1{12} a_i (a_i^2-1) (8a_j a_k-a_j^2-a_k^2)\, x^3 + {\cal O}(x^4)\,.
\ee
%%%
To ensure $\cosh\varphi_i \ge 1$, it is necessary that $a_i\ge 1$.
We verify numerically that smooth solutions exist for
$a_i\ge 1$, and that the functions $H_i$ behave as the following at
asymptotic infinity:
%%%%%
\be
H_i=1 + \fft{q_i}{x} -\fft{(c_1^i)^2}{2x^2} + \cdots\,.
\ee
%%%
Here $q_i$ is the charge parameter for the gauge fields $A_i$, while
$c_1^i$ is the coefficient of the normalizable scalar mode according
to
%%%%%
\be
\varphi_i= \fft{c_1^i+c_2^i\log x}{x} + \cdots
\ee
%%%%
at infinity.  In particular, the coefficient $c_2^i$ of the non-normalizable
mode always vanishes for these BPS bubbles.

The single-charge case admits an analytic solution of the form given
in (\ref{eq:bubsol}).  This gives the exact relation
$c_1=\sqrt{q_1(q_1+2)}$ highlighted in (\ref{normalization}). For the
generic multi-charge cases, we have been unable to attain an
analytical solution to (\ref{eq:newbubeqn}).  Nevertheless, we may
highlight some of the general features of any such solution.  To do
so, it is convenient to perform the transformation $\xi_i=x+\eta_i$.
The resulting differential equation is then
\begin{equation}
\eta_i''=-\eta_i'(\eta_i'+2)\left[x+\eta_i+\left(\fft{x}{x+\eta_j}\right)
\left(\fft{x}{x+\eta_k}\right)\right]^{-1}.
\label{eq:newbeq}
\end{equation}
The main purpose for introducing this transformation is to arrive at
the asymptotic forms
\begin{equation}
\eta_i\sim (a_i-1)x+\mathcal O(x^2)\quad\hbox{as}\quad x\to 0,\qquad
\eta_i\sim q_i+\mathcal O\left(\fft1x\right)\quad\hbox{as}\quad x\to\infty.
\label{eq:asyform}
\end{equation}
Note also that it is only the last two terms in (\ref{eq:newbeq}) which
couple the equations for the three fields together.

The AdS vacuum solution is obtained by taking the trivial solution
$\eta_i=0$.  Hence, for small deformations of AdS (corresponding to
small changes $q_i\ll1$), we may expect $\eta_i\approx0$.  More
precisely, by assuming $\eta_i\ll x$, the above equation may be
approximated by
\begin{equation}
\eta_i''=-\fft{\eta_i'(\eta_i'+2)}{x+1}.
\end{equation}
This approximation also has the feature that the three equations completely
decouple in this limit.  Furthermore, this now admits an exact solution
consistent with (\ref{eq:asyform}):
\begin{equation}
\eta_i=\lambda_i\log\fft{1+x/(1-\lambda_i)}{1+x/(1+\lambda_i)}.
\end{equation}
Of course, this is only consistent with our assumption $\eta_i\ll x$
for $\lambda_i\ll1$.  In this limit, we obtain the approximate solution
\begin{equation}
\eta_i\approx2\lambda_i^2\fft{x}{1+x}\qquad(\lambda_i\ll1).
\end{equation}
Asymptotically, we read off
\begin{equation}
q_i=2\lambda_i^2,\qquad c_1^i=2\lambda\qquad(\lambda_i\ll1),
\end{equation}
in which case we have demonstrated that
\begin{equation}
c_1^i\sim\sqrt{2q_i}\quad\hbox{for}\quad q_i\to0.
\end{equation}
Physically, we see that since small deformations decouple from each
other, we are allowed to turn on any independent combination of the
three commuting $R$-charges ($q_1$, $q_2$ and $q_3$) as we wish, while
maintaining the form of the 1/8 BPS bubble.

Larger deformations may be treated numerically.  However, it is worth
noting the general feature of (\ref{eq:newbeq}): since $\eta$
starts with a positive slope (forced by demanding
$\cosh\varphi_i\ge1$), it will remain positive with $\eta_i''\le0$.
This means $\eta$ is monotonic increasing with decreasing slope, and
will asymptotically approach its value at infinity,
$\eta_i(x\to\infty)=q_i$.  The initial slope for $\eta_i$, which is
$a_i-1$,then determines how large the final charge $q_i$ becomes;
larger $a_i$ gives larger $q_i$.

With this in mind, we may approximate (\ref{eq:newbeq}) by noting that
the expression $x/(x+\eta_i)$ is bounded to lie between $0$ and $1$.
This expression starts at its minimum value $1/a_i$ when $x=0$ and
increases towards $1$ as $x\to\infty$.  We now consider the
denominator in (\ref{eq:newbeq}):
\begin{equation}
\hbox{den}=x+\eta_i+\left(\fft{x}{x+\eta_j}\right)
\left(\fft{x}{x+\eta_k}\right).
\end{equation}
For small $x$, this is dominated by the initial value of the last
term, $1/(a_ja_k)$, while for large $x$ it is dominated by the first
term.  This suggests that we may make the approximation
\begin{equation}
\hbox{den}\approx x+\eta_i+\fft1{a_ja_k},
\label{eq:denapprox}
\end{equation}
in which case the second-order equation
\begin{equation}
\eta_i''=-\fft{\eta_i'(\eta_i'+2)}{x+\eta_i+(a_ja_k)^{-1}}
\end{equation}
admits a simple solution:
\begin{equation}
\eta_i=\sqrt{x^2+2(q_i+(a_ja_k)^{-1})x+(a_ja_k)^{-2}}-x-(a_ja_k)^{-1}.
\end{equation}
This is, in fact, exact for the one-charge case (where we take, {\it e.g.}
$a_2=a_3=1$ and find $\eta_1=\sqrt{x^2+2(q_1+1)x+1}-x-1$).  Here we have
chosen the constants of integration in accordance with the asymptotic
form (\ref{eq:asyform}). The $R$-charge is simply $q_i$, and the scalar
vevs are
\begin{equation}
c_1^i=\sqrt{q_i(q_i+2(a_ja_k)^{-1})}.
\label{eq:svevse}
\end{equation}

For these expressions to be self-consistent, we examine the behavior of
$\eta_i$ as $x\to0$:
\begin{equation}
\eta_i\sim q_ia_ja_kx+\mathcal O(x^2).
\end{equation}
Comparing this with (\ref{eq:asyform}), we see that self-consistency demands
\begin{equation}
a_i=q_ia_ja_k+1\qquad(i\ne j\ne k),
\end{equation}
which may also be expressed as
\begin{equation}
q_i=(a_i-1)a_j^{-1}a_k^{-1},
\label{eq:charres}
\end{equation}
where the initial slopes must satisfy $a_i\ge1$.  This condition gives
a range for the allowed values of the charges.  For example, with three
equal charges, the above reduces to
\begin{equation}
q=a^{-1}(1-a^{-1}),
\end{equation}
which has a maximum value $q_{\rm max}=1/4$, which occurs when $a=1/2$.

Of course, this restriction (\ref{eq:charres}) is entirely contingent
on the validity of the approximation (\ref{eq:denapprox}).
Numerically, we find that the above captures the qualitative behavior
of the solutions, but that the actual restrictions on the charges is
different.  In particular, numerically we find two possibilities for
the behavior of $q_i(a_i)$.  For one-charge and two equal charge
bubbles, $q_i$ can be made arbitrarily large.  However, for three
equal charge bubbles, the actual limiting value of the charge is given
by
\begin{equation}
q< q_{\rm max},\qquad q_{\rm max}\approx 0.529.
\end{equation}
This cutoff on the maximum amount of $R$-charge that can be supported
by a three equal charge BPS bubble may have implications on the
nature of the corresponding 1/8 BPS configuration in the dual
super Yang-Mills theory.

For the multi-charge cases where one of the $q_i$ can become large,
the scalar vev expression (\ref{eq:svevse}) applies, and we find
%%%%%
\be
\fft{c_1^i}{q_i} \rightarrow
\begin{cases}
\sqrt{2/q_i} &\hbox{for }q\rightarrow 0;\\
           1 &\hbox{for }q\rightarrow \infty.
\end{cases}
\ee
%%%%%
Even in regions where $q_i$ cannot become arbitrarily large, the dominant
behavior is for $c_1^i$ to approach $q_i$ from above as $q_i$ increases.
Similar results can also be obtained for $D=4$, $6$ and $7$ dimensions,
given by
%%%%%
\bea
D=4:&& H_i=a_i - \ft16 (a_i^2-1) a_j a_k a_\ell\, x^2 +
\ft13 a_i(a_i^2-1) a_j^2
a_k^2 a_\ell^2\, x^3 + {\cal O} (x^4)\,.\nn\\
D=6:&&
H_1=a_i-\ft9{10} (a_i^2-1) a_j x^\fft23 + {\cal O}(x^\fft43)\,,\nn\\
D=7:&&H_i = a_i -\ft23 (a_i^2-1) a_j\, x^\fft12 +
\ft16 a_i(a_i^2-1)(4a_j^2-1)\, x +{\cal O}(x^{\fft32})\,,
\eea
%%%
where the $i, j,k$ indices are not equal when they arise within the
same equation.  Using this as initial data near $x=0$, we can find
numerically that smooth BPS bubbles exist in all cases for $a_i\ge 1$.

\section{Conclusions}

We have investigated the non-BPS analog of `bubbling AdS' geometries
in type IIB supergravity, corresponding to a special class of non-zero
temperature LLM configurations. From the five-dimensional point of
view, these are solutions of the STU model coupled to three additional
scalars $\varphi_i$.  These solutions can be considered as the
bubbling generalizations of non-extremal AdS black holes, and have
regular horizons. However, unlike the previously-known AdS black
holes, the bubbling AdS geometries do not have a straightforward
generalization away from the BPS limit. Thus, we have had to rely on
various approximation methods, as well as numerical analysis. In
particular, we have considered the linearized $\varphi_i$ equations on
the background of the non-extremal $R$-charged AdS black hole. The
backreaction onto the other fields occurs at higher order. Even at the
linearized level, the equations do not have closed-form
solutions. Specifically, the linearized $\varphi_i$ equations can be
mapped to the Heun equation, for which the general two-point
connection problem remains unsolved.

Nevertheless, we can find approximate solutions to the Heun equation
by matching solutions in two overlapping regions. This method is
reliable for a certain regime provided that there is a large
overlap. In our case, the relevant regime is $T\gg\mu_i$, where
$T$ and $\mu_i$ are the temperature and chemical potentials of the
dual thermal Yang-Mills theory, respectively. If, in addition to this,
we consider a high temperature limit for which $T\gg 1$, then the $\varphi_i$ equations
can be solved without matching.  Corrections of the
order $\mu_i/T$ and $1/T$ can then be considered via a
perturbation approach.  The complete $\varphi_i$ solution is fixed by
the boundary conditions at the horizon. Namely, we require that
$\varphi_i$ is regular at the horizon. For the most part, we focus on
the case of one charge and a single additional scalar field
$\varphi_1$.

The behavior of $\varphi_1$ at asymptotic infinity can be related to
perturbations away from the UV superconformal fixed point of the
Yang-Mills theory. In particular, the normalizable mode of $\varphi_1$
goes as $1/r^2$ and corresponds to the dimension two chiral primary
operator $\Tr\,Z_1^2$ getting a vev. The non-normalizable mode goes as
$(\log r)/r^2$ and corresponds to a massive deformation of the field
theory Lagrangian of the form $\Tr\,Z_1^2$. We have normalized
$\varphi_1$ such that the vev is fixed at the value it has for the BPS
AdS bubble. Then we discuss how the thermal mass depends on the
physical parameters of the field theory, namely the temperature,
chemical potential and $R$-charge. We have considered both the grand
canonical ensemble and the canonical ensemble. The fact that the
$\Tr\,Z_1^2$ term is only present in the Lagrangian for thermal field
theories might indicate that there is a phase transition at zero
temperature. Moreover, we find the thermal mass vanishes at the point of the Hawking-Page transition. 

We have gone beyond the linear order in $\varphi_1$ to take into
account the backreaction on the other fields, In particular, we have
used this to obtain the asymptotic form of the corrections in order to
read off the mass of the bubbling AdS black hole. It would be interesting
to further consider the backreaction, or else use numerical analysis,
to investigate the thermodynamics of bubbling AdS black holes. In
particular, it would be interesting to find out how the local
thermodynamic stability constraints, as well as the Hawking-Page
transition, may be altered due to $\varphi_i$.

It would be useful to understand more concretely whether these
bubbling AdS black hole solutions have any relation to the hyperstar
solutions considered in \cite{buchel}. There, it was suggested that
the hyperstar background did not have a horizon because the
coarse-graining was taken over only the half-BPS sector of the full
Hilbert space of type IIB theory. Perhaps these bubbling AdS black holes
are the result of including the non-BPS sector in the coarse-graining.

Finally, we have shown via numerical analysis that there are actually
two types of non-BPS AdS bubbles. Thermal AdS bubbles, of which the
bubbling AdS black holes are a subset, have an event horizon
surrounding a singularity. On the other hand, solitonic AdS bubbles
are completely regular and horizon-free.  The latter type of non-BPS
bubbles can be obtained from the BPS bubbles by smooth deformations,
and therefore correspond to non-supersymmetric deformations of the
dual field theory. It would be interesting to investigate this
further.

\section*{Acknowledgments}

We should like to thank Wei Chen and Mirjam Cveti\v{c} for helpful conversations. The
research of H.L., C.N.P. and J.F.V.P. is supported in part by DOE
grant DE-FG03-95ER40917. The research of J.T.L. is supported in part
by the DOE grant DE-FG02-95ER40899.

%%%%%%%%%%%%%%%%%%%%%%%%%%%%%%%%%%%%%%%%

%%%%%%%%%%%%%%%%%%%%%%%%%%%%%%%%%%%%%%%%

%%%%%%%%%%%%%%%%%%%%%%%%%%%%%%%%%%%%%%%%
\end{document}